%% file: main.tex
\documentclass[manuscript,screen]{acmart}

\AtBeginDocument{%
  \providecommand\BibTeX{{%
    \normalfont B\kern-0.5em{\scshape i\kern-0.25em b}\kern-0.8em\TeX}}}

\setcopyright{acmcopyright}
\copyrightyear{2023}
\acmYear{2023}
\setcopyright{acmlicensed}\acmConference[FAccT '23]{2023 ACM Conference on Fairness, Accountability, and Transparency}{June 12--15, 2023}{USA} % {Chicago, IL, USA}
\acmBooktitle{2023 ACM Conference on Fairness, Accountability, and Transparency (FAccT '23), June 12--15, 2023, Chicago, IL, USA}
\acmDOI{10.1145/3593013.3594014}
\acmISBN{979-8-4007-0192-4/23/06}

\begin{document}

\title{Algorithms as Social-Ecological-Technological Systems: an Environmental Justice Lens on Algorithmic Audits}

\author{Bogdana Rakova}
\affiliation{%
  \institution{Mozilla Foundation}
  \city{San Francisco}
  \country{USA}}
\email{bogdana@mozillafoundation.org}

\author{Roel Dobbe}
\affiliation{%
  \institution{Delft University of Technology}
  \city{Delft}
  \country{Netherlands}}
\email{r.i.j.dobbe@tudelft.nl}

\renewcommand{\shortauthors}{Rakova and Dobbe}

\begin{abstract}
This paper reframes algorithmic systems as intimately connected to and part of social and ecological systems, and proposes a first-of-its-kind methodology for environmental justice-oriented algorithmic audits. How do we consider environmental and climate justice dimensions of the way algorithmic systems are designed, developed, and deployed? These impacts are inherently emergent and can only be understood and addressed at the level of relations between an algorithmic system and the social (including institutional) and ecological components of the broader ecosystem it operates in. As a result, we claim that in absence of an integral ontology for algorithmic systems, we cannot do justice to the emergent nature of broader environmental impacts of algorithmic systems and their underlying computational infrastructure. 
Furthermore, an integral lens provides many lessons from the history of environmental justice that are of relevance in current day struggles for algorithmic justice.
We propose to define algorithmic systems as ontologically indistinct from Social-Ecological-Technological Systems (SETS), framing emergent implications as couplings between social, ecological, and technical components of the broader fabric in which algorithms are integrated and operate. We draw upon prior work on SETS analysis as well as emerging themes in the literature and practices of Environmental Justice (EJ) to conceptualize and assess algorithmic impact. We then offer three policy recommendations to help establish a SETS-based EJ approach to algorithmic audits: (1) broaden the inputs and open-up the outputs of an audit, (2) enable meaningful access to redress, and (3) guarantee a place-based and relational approach to the process of evaluating impact. We operationalize these as a qualitative framework of questions for a spectrum of stakeholders. Doing so, this article aims to inspire stronger and more frequent interactions across policymakers, researchers, practitioners, civil society, and grassroots communities.
\end{abstract}

\maketitle

\section{Introduction}
\label{sec1}
% ==================== Introduction =================================
% Planetary boundaries as part of the motivation ... meaning we need to be cocnsiderate of downstream impacts, broader ecoclogical resources ... motivates an example of infrastructural AI - resource intensive 
% [THE INTRO CAN BE MADE MORE FOCUSED AND CONCRETE STRUCTURED TO MOTIVATE THE KEY CONTRIBUTIONS] 
Recent critiques of the use of data and algorithmic systems have shown the need for more comprehensive lenses that expand the analysis of impacts beyond the datasets and algorithms, situating algorithmic audits in a social and ecological context \cite{benjamin2019assessing, green2020algorithmic, dobbe2021hard}. The rapid rise in popularity of Artificial Intelligence (AI) has received related critiques pointing at the lack of considerations of social sciences and humanities \cite{sloane2019ai}, and deficits to address harmful outcomes of AI systems in the academic peer review processes \cite{hecht2021s}. For the purposes of this paper, we use the terms AI, algorithms, and AI systems interchangeably to refer to products or services that leverage automated decision-making processes. Throughout these prior critiques, there is a growing consensus to acknowledge the necessity of plurality in truths while dealing with a shared resource or system. While these critiques have recently emerged associated with the advent of new computing applications that lean on data, high-dimensional compute, and algorithms, they are not new. There have been earlier movements making comparable arguments, such as investigating a critical technical practice to reform AI in the 90s \cite{agre1997toward, malik2021critical}, the informatics of domination \cite{haraway2006cyborg}, and alternative modes of heterogeneity in technology \cite{doi:10.1111/j.1467-954X.1990.tb03347.x}. 

From a Science and Technology Studies (STS) perspective, the social and technical components of a system are inextricably connected and ontologically non-distinct \cite{bijker1997bicycles, jasanoff2015future}. Furthermore, scholars have brought attention to the constitutive ways in which algorithmic harms emerge from concrete choices made by algorithmic system developers, failing to account for the dynamic nature of social relations within which technical artifacts are situated \cite{green2020algorithmic, selbst2019fairness}. Through a comparative study of the history of impact assessment and algorithmic accountability, \citet{jake2021} highlight that "the impacts at the center of algorithmic impact assessments are constructs that act as proxies for the often conceptually distinct sociomaterial harms algorithmic systems may produce." Sociomaterial practices as in "the constitutive entanglement of the social and the material in everyday life" need to be understood in relation to technology and the organizational and political structures that produce particular technologies \cite{doi:10.1177/0170840607081138}. The disparate impacts and struggles within the practice and materiality of everyday life have been at the core of justice-focused studies of algorithmic harms \cite{noble2018algorithms, hanna2020towards, ogbonnaya2020critical, benjamin2020race, davis2021algorithmic}. Social justice civil society groups have been on the forefront of these investigations, and their research has been instrumental in formulating more comprehensive accounts of algorithmic harm \cite{costanza2020design, buolamwini2018gender, raji2019actionable}. 

In parallel, Environmental Justice (EJ) scholarship and grassroots communities have been leading actors in addressing harms to marginalized communities and the environment \cite{taylor2000rise, mohai2009environmental, pellow2000environmental}. Their struggles span over a longer time horizon and have led to effective strategies for more comprehensive, integral, and inclusive actions in response to environmental injustice. We leverage the evolution of interpretations, debates, and practical methodologies that have been core to the EJ field \cite{just2003, agyeman2016trends, holifield2001defining}, in broadening the abstraction boundaries around socio-technical AI systems. Given the more recent advent of algorithmic harms and the relative early phase of algorithmic justice as a field, we claim that there are many lessons that can carry over from the more mature EJ field. They could inform the lifecycle of algorithmic impact assessments while also contributing to building effective movements that challenge algorithmic harms and promote algorithmic justice. 

Recent work has engaged with the intersection of algorithmic systems and environmental sustainability \cite{dobbe2019ai, van2021sustainable, rohde2021sustainability} highlighting the lack of sufficient transparency into the energy, carbon, and broader environmental footprint of digital infrastructures and the tendency of data-heavy artificial intelligence techniques to exacerbate these, often in ways that reify existing injustices. 
However, to the best of our awareness, there is lack of studies investigating the implications of key EJ concerns within the AI impact assessment process. 
To fill this gap, we take inspiration from how the EJ field has approached the complexity of technologies and their social and ecological impact. Central in this approach is a system lens that brings into view the couplings and emergent dynamics between technological, social, and ecological elements of the broader system in which the technology resides \cite{leach2010dynamic}.
In what follows, we introduce such an approach to allow scholars, engineers, policy-makers, civil society organizers, and other diverse stakeholders to address these emergent challenges through understanding algorithmic systems as ontologically indistinct from Social-Ecological-Technological Systems (SETS). As such, a SETS approach becomes a foresight tool that enables us to expand the assumed boundaries of algorithmic systems and question their broader implications in a more-than-human world. Through bringing in perspectives from scholars studying social-ecological and social-ecological-technical systems \cite{ahlborg2019, biggs2021routledge, leach2010dynamic}, we aim to empower diverse stakeholders involved in the design and practice of algorithmic audits, to unpack and address a broad range of concerns.

The remainder of this paper is organized as follows: in Section \ref{sec2} we draw attention to emergent challenges in the algorithmic auditing ecosystem. We motivate a qualitative questions framework that integrates considerations from SETS and EJ, which we detail in Sections \ref{sec3} and \ref{sec4}, respectively. In conclusion, in Section \ref{sec5} we offer a starting point for diverse practitioners to situate themselves within the thematic challenges emerging from the field of EJ and work towards new practices for evaluating and overcoming harm and injustice in algorithmic social-ecological-technological systems.

% ============================================================

\section{Emergent challenges in the algorithmic auditing ecosystem}
\label{sec2}
% [LET'S INTEGRATE MOTIVATION INTO INTRODUCTION - I DON'T THINK WE NEED TOO MANY IDEAS AND INFLUENCES. BY DETERMINING THE CORE ISSUE, CORE PROBLEMS, WHAT IS MISSING AND OUR KEY CONTRIBUTIONS WE CAN SELECT THE MAIN INSPIRATION. NOW IT FEELS LIKE THERE ARE TOO MANY CONVOLUTED CONCEPTS THAT EXPLAIN THE SAME PHENOMENA IN SLIGHTLY DIFFERENT WAYS]

% Prior work in alg audits
A growing body of work across interdisciplinary fields has engaged with algorithmic auditing and impact assessment as a mechanism for algorithmic evaluation and accountability. In a recent field scan, \citet{costanza2022audits} conduct an anonymous survey among 189 individuals who lead algorithmic audits or whose work is directly relevant to algorithmic audits, identifying that their biggest challenges are lack of buy-in for conducting an audit and lack of enforcement capabilities. They also find that predominantly practitioners in the algorithmic auditing ecosystem rely on quantitative tools instead of qualitative methods, leveraging "quantitative ‘definitions’ of fairness to assess whether or not the distribution of goods and services across various subpopulations are equitable or fair" \cite{costanza2022audits}. This is a challenge that \citet{davis2021algorithmic} frame as \emph{algorithmic idealism}, "enacting computation that assumes a meritocratic society and seeks to neutralize demographic disparities." A dislocated sense of accountability~\cite{widder2022dislocated} and the lack of an accountability forum \cite{jake2021} contribute to tensions when practitioners engage in operationalizing algorithmic auditing frameworks in practice \cite{raji2020closing, rakova2021responsible}. % Furthermore, scholars have articulated issues of "scale thinking" as in prioritizing scalability. Building on work by anthropologist Anna Lowenhaupt Tsing \cite{tsing2012nonscalabilitythe}, \citet{hanna2020against} articulate scalability as "the ability of a system to expand without having to change itself in substantive ways or rethinking its constitutive elements". Furthermore, prioritizing scalability has implications for justice-oriented approaches to addressing algorithmic harms within transformative and restorative justice efforts \cite{hasinoff2022scalability}. In interviews with AI audit practitioners, \citet{rakova2021responsible} find that growing tensions exist in resisting scale thinking, while instead, scale could be achieved through partnership-based and multistakeholder frameworks. The challenge that interdisciplinary scholars identify is that the logics of scale at the core of the AI development lifecycle is antithetical to equity-driven models of justice \cite{hanna2020against, hasinoff2022scalability}.     

Another trend concerns growing critiques of the efficacy of audits and documentation-based practices for promoting safer and more equitable algorithmic systems.
\citet{cath_dutch_2021} argue that recent enthusiasm around algorithm registers in Europe deserves further scrutiny, as their empirical insights from digital welfare contexts show that the register does not situate algorithms in a sufficiently rich social and political context to properly understand its true social impact.
\citet{gansky_counterfacctual_2022} demonstrate that an emphasis on documenting and reflecting on contextual factors in the design and assessment of algorithmic systems abstracts away its reliance on institutions and structures of justice. They emphasize how justice infrastructures are often under-equipped to manage disputes, and how the current political economy of AI implementation, prioritizing profit seeking over duties of care, further obstructs realizing rights.

One crucial part of this political economy is laid out in detail by Whittaker, who explains how modern AI fundamentally depends on corporate data, compute resources, and business practices \cite{whittaker_steep_2021}. She warns about the growing reliance on AI in industry, consolidating power over our lives and institutions, while also using their financial and geopolitical power to determine much of what is possible to counter their power in terms of research, policy making, civil advocacy, and activism.
These politics have heavily shaped the discourse within AI research communities \cite{birhane_values_2022}, and are increasingly encroaching on research addressing the social and political impact of AI \cite{young_confronting_2022}, thereby preventing an honest and inclusive account of algorithmic harms and their political dimensions.

The same concentration of power and resources has been constitutive to another unavoidable trend: the ability to use computational resources to build large machine learning models.
In recent years, awareness has grown about the environmental impact of such algorithmic systems and their underlying computational infrastructures. \citet{strubell2019energy} estimate the environmental costs of Large Language Models (LLMs), while \citet{lottick2019energy} offer 14 policy recommendations on how to reduce it. Initiatives like Green AI \cite{schwartz2020green}, Green Deep Learning \cite{xu2021survey}, and online tools like MLCO\textsubscript{2} calculator\footnote{https://mlco2.github.io/}, CodeCarbon\footnote{https://codecarbon.io/}, and CarbonTracker\footnote{https://carbontracker.org/} aim to shift existing norms in AI development by advocating for the reporting of carbon footprint alongside accuracy and other efficiency metrics in AI. Increasingly, scholars have worked on estimating the carbon footprint of LLMs, spanning up to hundreds of billions of parameters and requiring millions of computing hours to train \cite{strubell2019energy, luccioni2022estimating}. Carbon footprint is only one dimension of environmental impact and in particular the couplings between environmental ecosystems and technical infrastructure \cite{dobbe2019ai}, which we frame as ecological-technological couplings in the next section. Other aspects of ecological-technological couplings issues, for example, water pollution due to cloud compute companies' operations \cite{waterjustice}, have received much less attention. 

Next, we point to social-technological couplings between LLMs and social norms, behaviors, and value systems which evolve as people interact with the technology. Within this aspect of downstream AI impacts, \citet{bender2021dangers} call for the consideration of how risks and benefits of LLMs are distributed, leveraging the literature on environmental justice and environmental racism \cite{westra2001faces, bullard1993confronting}. Furthermore \citet{bietti2019data} discuss how cloud computing companies, and Amazon in particular, contribute to indirect environmental harms through contracts with oil and gas industry as well as donations to political candidates denying climate change. 

Lastly, another dimension of LLMs' downstream impact we need to consider is related to social-ecological couplings. At its core, the field of social-ecological systems research engages with interactions between people and nature \cite{biggs2021routledge}. Scholars in this field ask questions related to knowledge co-creation and human decision-making which will inevitably be disrupted by the wide-scale use of LLMs in society. In sum, we aim to build on prior work arguing that increasing carbon emissions are undeniably coupled with algorithmic systems and deserve urgent attention. Furthermore, resisting the approach to equate environmental impact of AI to carbon footprint, we turn to prior work in the field of EJ.

Justice frameworks have been central to a growing number of investigations of algorithmic harms \cite{sloane2019inequality, costanza2020design, eubanks2018automating, noble2018algorithms}. 
We see the potential to further expand the bounds of algorithmic justice to both learn from and also take into account EJ concerns and benefit from its longer history. 
In this paper, we conceptualize EJ through a broad frame that includes theoretical scholarship, policy, and practice. The evolution of methodologies and multiple interpretations of EJ have contributed to emergent themes within the EJ frame and discourse, including: (1) focus on the practice and materiality of everyday life; (2) the impact of community, identity, and attachment; and (3) growing interest in the relationship between human and non-human assemblages \cite{agyeman2016trends}. In what follows, we show how EJ scholars have grappled with these issues and their relevance to calls for algorithmic justice. We aim to highlight and uplift non-Western, decolonial, and intersectional voices in how broadening the conception of the environment and EJ \cite{wald2019latinx, png2022tensions} could positively contribute to a just and equitable algorithmic auditing process in a more-than-human world.

% we argue that the ecological implications of algorithms and their computational infrastructures go far beyond, also including the use of scarce resources, impact on local water ecosystems, waste disposal, and accelerated fossil fuel production, to name a few ~\cite{dobbe2019ai, brevini2021ai_}. Hence, resisting the approach to equate environmental impact of AI to carbon footprint, we turn to prior work in the field of Social-Ecological-Technical Systems (SETS) analysis.

Algorithmic systems don’t exist in a vacuum but are technical infrastructure embedded in a social and ecological context \cite{leach2010dynamic}. \citet{ahlborg2019} provide an in-depth review of prior literature, interdisciplinary research, and empirical work, to show that “excluding technology or environment from the problem formulation, conceptual framework, and the analysis [methodological choices] precludes conclusions or policy recommendations that engage these at anything but a superficial level.” For example, one of the challenges they highlight is that socio-technical and socio-ecological systems scholars have differing approaches to how they define system boundaries and choose temporal, spatial, and quantitative scales of observation and measurement frameworks. \citet{ahlborg2012mismatch} have illustrated this through empirical examples in Nepalese forest management, using a mixed-methods approach in revealing the impact of disparate scales of knowledge. They highlight that “there are multiple positions within local knowledge systems and how these positions emerge through people’s use of and relations to the forest, in a dynamic interaction between the natural environment and relations of power such as gender, literacy, and caste." Furthermore, algorithmic systems used in forest ecosystem management have predominantly relied on a different scale of observations - satellite imagery and sensor data - which environmental scientists have argued is introducing temporal and spatial biases which haven’t been thoroughly investigated \cite{mcgovern2021need, rolnick2022tackling}. Without a broader and more integral lens addressing the couplings of algorithmic systems with socio-ecological and socio-technical systems, the algorithmic impact assessment process is bound to miss how technical interventions threaten to exacerbate existing power imbalances and yield harmful consequences. Therefore, in the next section we detail an integral analysis framework and its relevance to the algorithmic auditing process.

% For example, algorithmic models are used to assess the risk of lead contamination of the water infrastructure in individual homes and neighborhoods in Flint, Michigan, USA. To train the models, \citet{abernethy2016flint} utilize voluntary residential water tests, historical records, and city infrastructure data. In follow up research, \citet{chojnacki2017data} explore questions of self-selection in the residential testing program, examining which factors are linked to when and how frequently residents voluntarily sample their water. The findings from the research team are distributed to Flint’s citizens through a web and mobile application funded by Google. From a EJ perspective, it becomes critical to address the sampling bias of volunteer residential testing as well as residents’ differential access to the estimations made by the model. Low-income communities that don’t have Internet connection may inevitably not have access to vital data about the risk of lead contamination for their household. 

% LLMs impacts in terms of SETS? 

\section{Introducing Social-Ecological-Technological Systems analysis}
In order to properly understand how outcomes of a \emph{system} relying on AI and algorithmic technologies come about, we benefit from questioning and illuminating the elements that construct and constitute that system. 
A complex dynamical systems lens is central to most sustainability studies.
Here, we leverage the definition of a system by \citet{leach2010dynamic} - a system is "a particular configuration of dynamic interacting social, technological, and environmental elements." 
The authors argue that there’s a need to understand systems in terms of their structures and functions - \emph{what are their boundaries?; how are interactions across these boundaries constructed and constituted?; what are their functional outputs and their consecutive impacts?} 
By drawing from French philosopher Michael Foucault's theories centered on the relationship between power and knowledge \cite{foucault1991foucault}, they highlight the role of \emph{constitution}. To understand constitutive relations in a system, we must understand: (a) the role of power dynamics among actors who determine the system boundaries, (b) the system’s internal dynamics, and (c) interactions with what’s external to the system boundaries.

A central approach to studying complex dynamical systems and their constitutive relations is Social-Ecological-Technological Systems (SETS) analysis.
The SETS approach is centered on investigating the couplings between social, ecological, and technological dimensions of a system which are further defined as the social-behavioral, ecological-biophysical, and technological-infrastructure domains of the system \cite{pineda2021examining, mcphearson2021radical, markolf2018interdependent}. These domains are interdependent and interact with each other through three different relationships or couplings. \citet{mcphearson2021radical} define these relationships as follows (see also Table \ref{tab:sets-table}):
\begin{quote}
“(1) social–ecological couplings refer to human–nature or social–ecological relationships, feedbacks, and interactions, such as how urban nature provides ecosystem services to support human health and 
wellbeing or linkages between stewardship of urban green spaces and ecosystem change; (2) social–technological couplings refer to the ways in which technology and human social systems interact such as providing ability to communicate globally through social media or the dependence on technological infrastructure to facilitate dense human living in cities; and (3) ecological–technological couplings refer to the different ways in which climate and biophysical systems impact technology [and vice-versa] such as wildfires which cause power outages or rising temperatures driving increased energy use for cooling technology in buildings which in turn contributes to the urban heat island." \cite{mcphearson2021radical} 
\end{quote}
The SETS couplings framework has emerged through the works of scholars from the fields of Ecosystem Studies, Environmental Studies, Sustainability Science, Urban Ecology, and others. \citet{ahlborg2019} provide a comprehensive review of prior literature in these fields, synthesizing four arguments for the development of SETS analysis - (1) the need to better understand technological mediation of human–environment relationships, (2) the lack of a critical lens on the ambivalence of technology in respect to its impact, (3) the need to assess how agency and power are mediated through technology, and (4) the complexity of dynamics of change across scales of human impact on environmental ecosystems. 

% THEORY OF CHANGE
% By developing an understanding of algorithmic systems as SETS, practitioners could leverage an integrated systems approach to better understand the environmental footprint of their system as well as better diagnose and address forms of algorithmic harm and injustice. In this work, it is not our intention to critique the challenges or approaches within a socio-ecological or a socio-technical frame of conceptualizing algorithmic impact. Instead, our \emph{theory of change} is centered on the need and utility of interdisciplinary collaborations in disentangling the complex nature of algorithmic impacts. We argue that the SETS approach and prior scholarship could inspire co-constitutive strategies that empower diverse practitioners to address emerging challenges. % Furthermore, the historical account of the struggles within communities at the forefront of EJ could more broadly inform the work on algorithmic audits.

\input{tables/sets_table.tex}
% [Better clarify the info in the table]

\label{sec3}
\input{tables/sets_vs_justice.tex}

By bringing in interdisciplinary prior work by scholars investigating socio-technical and socio-ecological systems, here, we motivate the use of a SETS approach to better understand the environmental dimensions of an algorithmic audit. We propose a set of questions that aim to help practitioners operationalize a SETS approach to the algorithmic audit. The questions we lay out in Table \ref{tab:sets-justice-table} serve as a starting point to enable further interdisciplinary dialogue and empirical investigations during an AI audit. We seek to make our analysis actionable for diverse stakeholders who would be best positioned to explore the EJ dimensions of algorithmic SETS impacts. The goal of the framework is to propose a SETS-based EJ approach to algorithmic audits, which we motivate through learning from the way the EJ community has grappled with issues similar to those within  the algorithmic justice field. The questions along the dimensions of Table \ref{tab:sets-justice-table} aim to integrate (1) the structural dimensions of a SETS analysis and (2) the emerging themes within EJ practice and scholarship. In the rest of this section, we outline key considerations and challenges in the social–ecological, ecological–technological, and social–technological couplings of a SETS analysis. Specifically, we focus our discussion on aspects which have not received much attention in the context of algorithmic systems and provide pointers to more literature for interested practitioners to further engage with.

% We recognize that the framework is a starting point and future work will need to offer a process and concrete empirical case studies. Towards that goal, in what follows we identify how scholars within SETS have grappled with these questions and draw policy recommendations that we believe could positively contribute to the algorithmic auditing ecosystem. % These are: (1) broadening the inputs and opening-up the outputs of an impact assessment, (2) de-centering power structures and enabling meaningful access to redress, and (3) adopting a place-based and relational approach to the process of evaluating impact.
   
\subsection{Social-ecological couplings}
\label{sec3.1}
The research within sustainability and sustainability transitions has been broadly influenced by the study of social and ecological systems linkages with regards to the complex interrelationships across human actors, natural ecosystems, and the role of institutions in decision-making and governance \cite{clark2020sustainability}. \citet{ostrom2007diagnostic} proposes a foundational Social-Ecological Systems (SESs) framework which investigates interactions between "(i) a resource system (e.g., fishery, lake, grazing area), (ii) the resource units generated by that system (e.g., fish, water, fodder), (iii) the users of that system, and (iv) the governance system." In particular, they point to the importance of analysis methodologies that establish a common vocabulary for the comparison of specific empirical settings. In their early work, SES scholars point to the use of SES frameworks and methodologies in investigating the questions of governance of technological systems such as power grids or telecommunications systems \cite{mcginnis2014social}. Furthermore, the SES body of research and practice has also developed a deep understanding of harms through the lens of vulnerability analysis \cite{turner2003framework, berrouet2018vulnerability}, which has emerged through the body of work on environmental risks and hazards \cite{cutter2002american, tobin1997natural}. Building on critical reflections on the ontological grounding of SES scholarship and drawing from the field of EJ, \citet{pineda2021examining} point out that a number of recent SETS studies have “only partially represented socio-ecological system interactions and often neglecting the technological infrastructure subsystems." To remedy that, they bring perspectives from the field of EJ theory and practice to propose and empirically test an evaluation framework grounded in qualitative questions that inform the selection of SETS measurement indicators \cite{pineda2021examining}. 

% \citet{mcginnis2014social} propose a foundational Socio-Ecological System (SES) framework which enabled "researchers from diverse disciplinary backgrounds working on different resource sectors in disparate geographic areas, biophysical conditions, and temporal domains to share a common vocabulary for the construction and testing of alternative theories and models that determine which influences on processes and outcomes are especially critical in specific empirical settings." 

\subsection{Ecological-technological couplings}
\label{sec3.2}
SETS scholars have argued that technological-infrastructure systems mainly "extract resources, pollute and destroy in their sources, through their flows, and in the places in which they end up as waste" \cite{pineda2021examining}, calling for the need to investigate ecological-technological couplings. A number of studies have been done investigating these interrelations in the context of the built environment \cite{anderies2014embedding, mcphearson2016advancing, rozzi2015earth}. Furthermore, a number of scholars in the socio-technical research domain have engaged in SES analysis in evolving governance frameworks that encompass technical infrastructure \cite{foxon2009governing, smith2010politics}.

In the context of AI, \citet{matus2022certification} argue that measuring algorithmic impact should build on analysis from the sustainability domain, specifically "the study of socioenvironmental issues and sustainability more broadly in commodities and supply chains." Learning from sustainability, they propose the use of a robust certification system, including: (1) a set of technical standards which reflects a theory of change, (2) a certification process which facilitates monitoring and enforcement, and (3) a labeling program to provide information to consumers. One of the major challenges they discuss is that existing measurement frameworks are driven by network effects, making consensus among diverse stakeholders an intrinsic rather than instrumental goal. 

Furthermore, we identify a challenge in terms of the ability of existing measurement frameworks to assess downstreams impacts of the use of an algorithmic system in a particular local context \cite{clutton2021climate, musikanski2018ieee}. While we have seen a growing number of scholarship with regards to measuring the carbon footprint of algorithmic systems, there has been less attention given to other environmental sustainability concerns such as land use, water pollution, deforestation, biodiversity, habitat destruction, over-exploitation of resources, and others \cite{dryer_ai_2020}. 

\subsection{Social-technological couplings}
\label{sec3.3}
At the core of the socio-technological couplings within a SETS approach are considerations related to the interactions between social and technical systems and their interrelationships with evolving social norms, behaviors, values, and belief systems. Many recent studies have pointed to the importance of understanding algorithmic systems as socio-technical systems i.e. a combination of social and technical components interacting in complex dynamic ways. By drawing on studies of socio-technical systems in STS, \citet{selbst2019fairness} introduce a critical framework that informs the need to contextualize machine learning systems in a way that expands their boundaries to include social actors rather than purely technical ones. Furthermore, a growing number of STS scholars have argued for the need to reconceptualize algorithmic harms within a theoretical model of justice. \citet{sloane2019inequality} argues that there’s a need to reconfigure algorithmic discrimination as an issue of social inequality by centering social justice in technology policy and computer science education. Similarly, in examining the freedom and justice dimensions of AI, \citet{coeckelbergh2021ai} suggests that “institutionally, we also need more permanent interfaces between, on the one hand, technology development (in industry, academia, etc.) and, on the other hand, political and societal discussions." We see that a justice oriented STS approach to addressing the impact of an algorithmic system through the lens of socio-technological couplings has already been adopted by a number of practitioners.
However, it needs to be mentioned that addressing the political dimensions of AI/algorithmic systems, whether in academia, among workers, policy makers or civil society, is challenged by the fact that most financial resources as well as the AI technologies and infrastructures themselves are increasingly controlled by a small number of tech companies \cite{whittaker_steep_2021}. This forms an inevitable socio-technical coupling and barrier that requires broader forms of organizing to overcome, and which we hope can be inspired by traditions in the EJ movement and related thinking promoted in this manuscript.

% Furthermore, scholars have argued for a deeper understanding of  systems’ social and political context. Leveraging participatory design and action research, scholars have also called for a relational approach to understanding algorithmic harms, “recognizing knowledge as partial and situated" \cite{katell2020toward}. Knowledge systems are connected to particular territories and ecologies. For example, traditional ecological knowledge encapsulates indigenous cultural practices, world views, and epistemological and ontological knowledge embedded in geographic representations \cite{berkes2000sacred}. With regards to AI, there have been a number of critiques of the dominant approach to the relationship between data and knowledge in the life cycle of algorithmic systems, calling for a reconfiguration of power structures \cite{mhlambi2020rationality, rakova2020leveraging}.

\section{Mirroring Algorithmic Justice to Environmental Justice concerns}
\label{sec4}
The explored couplings necessary to understand algorithmic harm and injustice motivate the use of a SETS approach. However, it would be inefficient to apply the approach blindly, as the question remains what \emph{model of justice} to embrace to orient ourselves to the salient harms and vulnerabilities that need to be unpacked. Here we conceptualize a model of justice through lessons learned within the field of Environmental Justice (EJ). Theory and practice in the EJ movement has long grappled with struggles related to (1) distributive justice - equity in the distribution of environmental risk; (2) recognitional justice - recognition of the diversity of participants and experiences in affected communities; and (3) procedural justice - opportunities for participation in the political processes that create and manage environmental policy \cite{holifield2001defining}. As such, its legacy serves as a model of justice, both in its normative orientation as well as in its evolution as a field and a movement. In the most recent EJ annual review, \citet{agyeman2016trends} provide a comprehensive overview of EJ literature and goals in advancing climate solutions that link human rights and development in a human-centered approach, placing the needs, voices, and leadership of those who are most impacted at the center. 

% Justice frameworks have been central to a growing number of investigations of algorithmic harms\cite{sloane2019inequality, costanza2020design, eubanks2018automating, noble2018algorithms}. We see the potential to further expand the bounds of algorithmic justice to both learn from and also take into account EJ concerns. The evolution of methodologies and multiple interpretations of EJ have contributed to emergent themes within the EJ frame and discourse, including: (1) focus on the practice and materiality of everyday life; (2) the impact of community, identity, and attachment; and (3) growing interest in the relationship between human and non-human assemblages \cite{agyeman2016trends}. In what follows we show how EJ scholars have grappled with these issues and their relevance to calls for algorithmic justice. Furthermore, we aim to highlight and uplift non-Western, decolonial, and intersectional voices in how broadening the conception of the environment and EJ \cite{wald2019latinx, png2022tensions} could positively contribute to a just and equitable algorithmic auditing process in a more-than-human world.   
% Within each thematic area, we turn attention to related prior work on the impact assessment of algorithmic systems while highlighting potential gaps. In the following \emph{Section 5} we would adopt a SETS frame in search of a co-constitutive and integrated approach to addressing these gaps.

\subsection{Practice and materiality}
\label{sec4.1}
This theme within EJ encompasses (1) the politics of sustainability of everyday life, in the way we provide for basic human needs, for example, access to water, food, and energy, and (2) the unsustainable material flows that contribute to inequalities with regards to the access, security, and sovereignty to such resources on local, national, and global scales \cite{agyeman2016trends}. EJ concerns here are deeply intertwined with the issues of toxic waste, health disparities, food justice, indigenous justice, and climate justice. The focus on the materiality of everyday life also emphasizes the embodied experiences of environmental injustices by the human bodies of women, children, people of color, indigenous people, and marginalized communities. From an EJ perspective, “the material life of the body is seen as fully entangled with questions of social justice, providing a stronger foundation for an intersectional politics of alliance building between environmental and EJ movements; between the global north and south; and across the racial, cultural, sexual, and gender boundaries that mark bodies" \cite{agyeman2016trends}.  

% A growing number of scholars have documented the practice and materiality of algorithmic systems. 
% Exploitative and extractive data collection and use practices have been central actors within emerging theories of data justice, data capitalism, data colonialism, and data sovereignty, informing theories of algorithmic reparation, algorithmic coloniality, algorithms of oppression, and others \cite{thatcher2016data, kukutai2016indigenous, mohamed2020decolonial, noble2018algorithms, davis2021algorithmic}. Similarly to emerging trends in EJ, STS scholars have taken an intersectional approach to illuminating the power dynamics, structural conditions, and socio-technical context within which algorithmic systems exist \cite{hoffmann2019fairness, mhlambi2020rationality, katell2020toward}.

% Therefore, we bring attention to the need to incorporate concerns related to practice and materiality along the lifecycle of algorithmic systems and not only on the impact assessment stage. By operationalizing broader metrics and optimization frameworks, diverse stakeholders could also account for the resources that were utilized during the system development as well as the dynamic ways in which algorithmic outputs and the reliance on the underlying computational infrastructures may negatively impact social and environmental systems. 

In bringing awareness to the material flows in the lifecycle of an algorithmic system, practitioners could learn from the work of a growing number of communities at the forefront of the EJ movement. It is vital to learn about and acknowledge the historical, cultural, and sociopolitical context out of which specific grassroot communities have emerged and seek to understand if there could be better alignment and reciprocal relationships between (1) EJ efforts related to the practice and materiality of everyday life and (2) the critical work on investigating how material flows are impacted by algorithmic systems.  Exploitative and extractive data collection and use practices have been central actors within emerging theories of data justice, data capitalism, data colonialism, and data sovereignty, informing theories of algorithmic reparation, algorithmic coloniality, algorithms of oppression, and others \cite{thatcher2016data, kukutai2016indigenous, mohamed2020decolonial, noble2018algorithms, davis2021algorithmic}. Similarly to emerging trends in EJ, STS scholars have taken an intersectional approach to illuminating the power dynamics, structural conditions, and socio-technical context within which algorithmic systems exist \cite{hoffmann2019fairness, mhlambi2020rationality, katell2020toward}. 

Thus, an EJ approach to the algorithmic impact assessment process would involve the consideration of material resource flows along the entire lifecycle of an algorithmic system including the supply chains of the digital infrastructure on which it runs \cite{pasquinelli2020nooscope, widder2022dislocated}. Learning from and building meaningful relationships with existing civil society actors, grassroots movements, and local communities, there’s a need to understand how algorithmic systems disrupt these flows \cite{ganesh2022resistance, hasinoff2022scalability}, leading to higher order impacts to the livelihoods of people and the resiliency of environmental ecosystems. This is not an obvious and mechanical exercise and requires building respectful relationships that enable shared vocabulary. We highlight the need for the AI audit process to embrace a continuous self-reflexive practice, acknowledging the positionality of different actors involved, and how it shapes power imbalances and challenges towards mutual understanding reciprocity (see also Table \ref{tab:sets-justice-table}).

\subsection{Community, identity, and attachment}
\label{sec4.2}
The main argument in this body of EJ practice and scholarship is that environmental injustices have spatial and cultural dimensions. Major issues highlighted by \citet{agyeman2016trends} are related to the interdependencies between EJ and identity, community, space, place, and agency, linking to critiquing the problems of gentrification, displacement, and homelessnes. Furthermore, from the perspective of EJ, sustainability is seen as context-specific and relational. We find that a large number of prior work on algorithmic auditing and algorithmic harms investigations fall within this thematic area of EJ concerns \cite{noble2018algorithms, benjamin2019assessing}. Building on that we aim to highlight emerging learnings and synergies.   

From an EJ perspective, it is important to consider the way individual and collective identity is deeply connected to people’s attachment and connection to place \cite{anguelovski2014neighborhood}. While scholars have recently exposed how identity is shaped by algorithmic systems \cite{hanna2020towards, stark2019facial, keyes2021truth}, we see a need to further disentangle the relationships between identity, place, attachment, and algorithmic systems. Individual and collective identities are dynamic and mutable - they change and evolve over time and that process allows us to develop a more coherent sense of self. Diverse scholars have shown how algorithmic systems interfere with identity transitions - our ability to adapt to different life situations, such as gender transitions, divorce, career change, and others \cite{haimson2016constructing, haimson2015online}. Displacement due to environmental disasters inevitably leads to identity transitions for impacted communities and entire populations who might turn to AI-enabled products and services in crisis situations. While there have been analyses of the use of AI in the field of disaster risk management \cite{moitra2022ai, kuglitsch2022artificial}, we point to the need to more thoroughly account for a broad number of concerns emerging within EJ and algorithmic justice critiques. In sum, algorithmic audits need to engage with EJ-related data and situated knowledges within the communities where a particular AI technology is deployed \cite{kaack2022aligning}. Tracing the links between historical, representational, measurement, and deployment biases, needs to be grounded in and informed by the lived experiences of people impacted by the technology in a particular place.

\subsection{Human and non-human assemblages}
\label{sec4.3}
Finally, we highlight that an emerging theme within EJ discourses brings attention to the relationship between human and non-human sustainability. Key issues discussed by \citet{agyeman2016trends} encompass the environmental degradation of ecosystems, leading to increased vulnerabilities for their inhabitants. Fundamentally, EJ adopts a pluralist approach to the relationship between environment and justice - “an internal pluralism, incorporates a diversity of necessary forms of justice, rather than privileging only one, and retains flexibility in how functionings and flourishings are to be secured" \cite{walker2009beyond}.

EJ asks us to learn from the body of work in Ecolinguistics which is situated at the intersection of the fields of Human Ecology and Linguistics. It investigates \emph{the stories we live by}, postulating that stories are linguistic devices that we could analyze to judge their impact \cite{stibbe2015ecolinguistics}. Within the view of AI as an instrument of knowledge extractivism \cite{pasquinelli2020nooscope} or AI as an enabler for respectful and reciprocal non-human kin \cite{lewis2018making}, or other framings, we argue that a self-reflexive approach to the use of metaphors could enable improved outcomes in engaging with the social and environmental impacts of algorithmic systems. In alignment with prior work by STS scholars, we question the metaphors used in AI. In their literature review, \citet{stark2019data} demonstrate that “the metaphors we deploy to make sense of new tools and technologies serve the dual purpose of highlighting the novel by reference to familiar, while also obscuring or abstracting away from some features of a given technology or practice.” For example, \citet{seaver2019captivating} conceptualizes recommender systems as \emph{traps}, arguing that: 
\begin{quote}
"returning to the time structure of trapping makes the continuity between traps and [recommender system] infrastructures more visible: an infrastructure is a trap in slow motion. Slowed down and spread out, we can see how traps are not just devices of momentary violence, but agents of ‘environmentalization’, making worlds for the entities they trap." \cite{seaver2019captivating} \end{quote} 
Furthermore, diverse scholars have questioned who is the human in human-centered algorithmic systems \cite{chancellor2019human, aragon2016developing} calling for the account for non-human living beings \cite{umbrello2021ecological, ziesche2021ai, hg2020argument}. Building on their analysis we take inspiration from the field of EJ to engage practitioners in evolving algorithmic impact assessment frameworks that embrace a pluralist approach to considering the non-human in the co-constitution of algorithmic impact. In the next section we explore this further by leveraging a critical SETS approach to the algorithmic auditing process.

% In summary, through the lens of EJ, we see emerging the potential and need for the algorithmic impact assessment process to consider (1) the non-human as in non-human living beings and ecosystems and (2) nature’s rights as in environmental personhood and legal standing \cite{reeves2021responding, stone1972should}. In the next section we explore this further by leveraging a critical SETS approach to the algorithmic auditing process.

% ===================================================

\section{Discussion and Policy Recommendations}
% Integrating lessons from SETS and EJ in algorithmic auditing:
\label{sec5}
Adopting a complex SETS approach to the algorithmic auditing process helps in illuminating the systemic complexity of justice concerns, which allows for more effective normative debate and design of interventions. Algorithmic systems are ontollogically indistinct from SETS, that is: (1) they are dependent on social and ecological resources to be built, used, and maintained; (2) they contribute to carbon and ecological costs  through their dependency on computational infrastructure and their role in fossil fuel production; and (3) any problem that is solved through AI has underlying technical, social, and institutional components and their interactions and infrastructures depend on and impact SETS. In this section, we aim to further integrate learnings from SETS analysis and EJ practice, to give interdisciplinary practitioners a map of possible starting points to leverage them in the context of AI. The discussion that follows engages with the questions we discussed earlier in Table \ref{tab:sets-justice-table} - mapping considerations which need to be integrated in an iterative algorithmic auditing process along the lifestyle of an AI system. Towards that goal, we identify how scholars within SETS have grappled with these questions and draw policy recommendations that we believe could positively contribute to the algorithmic auditing ecosystem. These are: (1) broadening the inputs and opening-up the outputs of an impact assessment, (2) enabling meaningful access to redress, and (3) adopting a place-based and relational approach to the process of evaluating impact.

\subsection{Broadening the inputs and opening-up the outputs of an impact assessment}
\label{sec5.1}
% In the field of critical algorithmic studies, there have been a number of critiques of recently proposed algorithmic impact assessment frameworks which have come out of research within the consumer technology companies themselves \cite{bietti2020ethics, matus2022certification, costanza2022audits, raji2022outsider}. Predominantly, it is unclear if such frameworks are considering broad or narrow scope of inputs in terms of data and scenarios used for the evaluation of impacts. Similarly, there’s lack of transparency with regards to the auditing methodology and findings which we frame here as outputs of the assessment. For example, there have been recent controversies in algorithmic audits of candidate-screening algorithms used in hiring, which have been criticized for misrepresenting results, co-opting the audit outputs, and placing them entirely behind non-disclosure agreements \cite{hireview_2021, sloane2022silicon}.  

% Learning from SETS, we argue that there’s a need for methodologies that enable improved reflexivity by broadening out the inputs to algorithmic impact assessments and opening up their outputs. 

% We turn to methodologies and strategies within SETS analysis and discuss their relevance to improved transparency in algorithmic auditing practices. 
Scholars in SETS have taken a critical approach to integrating lessons between the fields of socio-ecological systems and socio-technical systems in the context of sustainability transitions \cite{leach2010dynamic, stirling2008opening}. Specifically, by addressing the questions of sustainability governance through a process of appraisal. The appraisal process is characterized by a particular methodology, the range or breadth-narrowness of inputs that are included (such as issues, perspectives, scenarios), and the range of outputs that are produced and conveyed to a wider audience through a closing-down or opening-up approach. For example, while a \emph{closed down} appraisal could result in unitary prescriptive policy recommendations, a focus on \emph{opening up} the appraisal “focuses on neglected issues, includes marginalized perspectives, triangulates contending knowledges, tests sensitivities to different methods, considers ignored uncertainties, examines different possibilities, and highlights new options” \cite{stirling2008opening}. 

% Sustainability appraisal is the process of assessment of policy, projects, and decision-making in sustainable development, which integrates analysis across environmental, social, and economic pillars of sustainability \cite{dalal2014sustainability}. Furthermore, \citet{stirling2008opening} frames the social appraisal of technology as “the ways in which knowledges, understandings, and evaluations are constructed and rendered salient to inform these commitments [commitments informing tangible social choices in the governance of technology].” \citet{leach2010dynamic} have argued for the need for appraisal methods to engage with the technical infrastructure dimensions of decision-making and the governance context within which they are embedded. 

Through a SETS lens, \citet{leach2010dynamic} call for moving away from narrow and closed risk-based approaches such as cost-benefit analysis and risk assessment, towards broad and open framings such as participatory and deliberative mapping. For example, \citet{burgess2007deliberative} point to the lack of mechanisms that enable contestation in prior methodologies and develop deliberative mapping as an approach engaging specialists and citizens in an appraisal process combining the use of quantitative decision analysis techniques with participatory deliberation. They use the methodology in managing the disposal of UK’s radioactive waste which is a fundamental struggle in the EJ field \cite{burgess2004citizens}. Following the evolution of SETS analysis and EJ concerns, we see that participatory methods have been an active area of debate in issues regarding climate change policy. In recent work, \citet{su14084656} provide an overview of different strategies and develop a framework for participatory value evaluation. Similarly, in algorithmic auditing, we have seen a growing number of calls for adopting participatory methods that enable meaningful community participation in the auditing process as well as disclosure of audit findings \cite{costanza2022audits, sloane2020participation, katell2020toward,shen2021everyday}. We hope that by drawing parallels in how these questions have been evolved in other policy domains, AI audit practitioners can leverage key lessons and opportunities for collaboration.  

\subsection{Enabling meaningful access to redress}
% De-centering power structures by enabling easier access to contestability
% De-centering to cede power in frontline communities
\label{sec5.2}
Informed by a SETS-oriented approach and recent EJ discourse, the algorithmic auditing process needs to enable meaningful access to redress of algorithmic harms. In practice, these could be operationalized through (1) prototyping mechanisms for reporting of algorithmic incidents, controversies, or injustice experienced by (un)intended users of AI systems, (2) considering how such mechanisms are affected by and may contribute to existing power structures, and finally (3) the role of workers in a growing number of calls for solidarity. 

Within EJ, a growing number of community-led initiatives examine how environmental data about harms and injustices moves between communities and policymakers \cite{eric2021, nost2021visualizing}. Recent developments in EJ have called for critical and decolonial frameworks, recognizing "power-dynamics and the complexity of intersectionality of marginalizations and injustices" \cite{menton2020environmental}, suggesting that "rather than simply participation, justice must include self-governing authority" \cite{temper2019blocking}. An algorithmic auditing ecosystem that empowers (un)intended users to contribute to improved understanding and transparency of potential algorithmic harms could lead to improved algorithmic accountability. However, there is a need to enable that through requiring meaningful avenues for users to report, challenge, contest, and seek redress for experiences of algorithmic harms. Furthermore, AI auditors conducting an assessment of a particular algorithmic system would greatly benefit from individual or collective accounts of experiences of algorithmic bias, harms, or injustice. Currently such incidents reporting data has predominantly been leveraged in third-party algorithmic audits lead by independent researchers with no contractual obligation to the audit target \cite{costanza2022audits}. \citet{shen2021everyday} conduct a field scan of examples where users have meaningfully contributed to third-party audits across particular AI systems including Google Image Search, Google Maps, YouTube, TikTok, and others. Learning from the field of EJ, we identify the need to create policy and other mechanisms that provide avenues for standardizing the participation of users in algorithmic auditing efforts. 

When engaging end users in an algorithmic audit, we need to consider how could this potentially contribute to existing power asymmetries and center-margin dynamics between AI companies and impacted intended and unintended users of their technology. The power-dynamics and justice aspects of participation have already been at the core of decolonial critiques of the design, development, and use of algorithmic systems \cite{agre1997toward, berg1998politics, ehsan2020human}. \citet{mohamed2020decolonial} explore the role of decolonial theory in critical analysis of the center-margin dynamics of power structures within an algorithmic system. Similarly, \citet{ricaurte2019data} incorporates a multidimensional approach to enable decolonial, intersectional, and feminist analysis of data colonialism in deconstructing algorithmic violence and empowering resistance. Furthermore, \citet{sloane2020participation} go on to critically examine existing modes of participation in the design and development of machine learning systems, cautioning against “participation washing”, that may legitimize exploitative and extractive forms of community involvement.

Power structures in AI are also being examined from within technology companies developing large scale AI systems. There is a growing movement and literature surrounding the role of workers in improving the understanding and awareness of algorithmic harms \cite{nedzhvetskaya2022role}. Within the EJ field, we have seen initiatives led by employees of technology companies including Amazon\footnote{https://amazonemployees4climatejustice.medium.com/amazon-employees-are-joining-the-global-climate-walkout-9-20-9bfa4cbb1ce3} and Microsoft\footnote{https://github.com/MSworkers/for.ClimateAction }, petitions banning the use of AI for fossil fuel exploration and development\footnote{https://onezero.medium.com/google-says-it-will-not-build-custom-a-i-for-oil-and-gas-extraction-72d1f71f42c8}, and calls for action from civil society organizations. \citet{dobbe2019ai} illuminate the interrelationships across many of these efforts in highlighting the role of tech workers in global EJ movements. Furthermore, they argue for acknowledging AI companies' contribution to the climate crisis and discuss recommendations for policymakers including: (1) mandating transparency, (2) accounting for the impacts of the supply chain of AI and its computational infrastructure \cite{crawford2018anatomy, widder2022dislocated}, (3) going beyond efficiency and being wary of rebound effects \cite{coulombel2019substantial}, (4) considering how AI is integrated in non-energy policy domains, (5) integrating tech regulation and green new deal policy making, (6) curbing the use of AI to accelerate fossil fuel extraction, and (7) addressing the use of AI to harm and exclude climate refugees. Inspired by an EJ frame, we hope policy guardrails in AI could leverage tactics evolved within the EJ movement, both in terms of enabling tech workers whistleblower protection as well as creating pathways for contestability and redress of algorithmic harms among communities disproportionately impacted by AI systems.   

\subsection{Adopting a place-based and relational approach to the process of evaluating impact}
\label{sec5.3}
% why is that important to alg justice? 
% what do the terms mean - a place-based and relational approach ?
Finally, we discuss place-based and relational approaches to SETS analysis and highlight their potential impact in the context of algorithmic auditing. A place-based approach or place-based policy commonly assumes the interactions between an intervention and a specific physical geography are critical in the implementation and impact assessment of that intervention \cite{barca2012case}. SETS scholars have demonstrated the role of place-based research methods as a key contributor to achieving global sustainability goals \cite{moran2010environmental}. For example, \citet{balvanera2017interconnected} argue that “transformations towards sustainability are often triggered at the local scale through the co-construction of local solutions." By analyzing the strengths and challenges to place-based socio-ecological research, they highlight that a place-based approach enables the consideration of long-term time horizons, capacity building, local communities of practice, and a bridge between local and global sustainability initiatives. Furthermore, SETS scholars are accustomed to studying systems in relational terms \cite{nkhata2008resilient, lejano2019relationality, west2020relational}. Recent work has sought to enable practitioners to operationalize relationality through describing and modeling not just material but also social processes that drive or impede resilience in a socio-ecological system, for example in disaster response and recovery \cite{cagney2016social} and urban planning \cite{goldstein2015narrating}.  

A growing number of interdisciplinary scholars have articulated a relational approach in understanding algorithmic impacts \cite{birhane2021algorithmic, viljoen2021relational, cooper2022accountability}. Furthermore, through a SETS lens, we argue that \emph{place} is where relational concerns need to be acknowledged, documented, debated, and ultimately addressed through diverse models of engagement. Being able to relate to environmental sustainability is often made possible through the place-based work of grassroots communities and civil society actors who bring awareness to the socio-economic and socio-ecological hardship and stressors for people and non-human living beings and ecosystems. Similarly, investigative journalists and civil society initiatives have often been at the forefront of algorithmic justice issues experienced by marginalized communities. 

% A vast number of scholars in the fields of Participatory Design, Action Research, Community Informatics, Value Sensitive Design, Structured Dialogic Design, Critical Studies, and others, have studied the complex social-ecological-technical challenges with regards to the lifecycle of technical interventions. Yet, we argue that applied, intersectional, and interdisciplinary future work is needed in meaningfully integrating participatory models of engagement particularly in the algorithmic impact assessment process. 

% do we see the same happening for algorithmic justice? Thinking of the many NGOs and activists?

% The practice of SETS analysis is context specific, place-based, and embedded in a complex dynamic socio-ecological context \cite{moran2010environmental}. 

What would it look like for policy makers and AI developers to enable local communities where the technology is deployed to evolve their own tools and processes for algorithmic audits? Learning from climate policymaking, we could leverage existing models for citizen deliberation and online participation contributing to a more meaningful algorithmic auditing process. For example, \citet{itten2022digital} present a case study from the Netherlands showing the benefits of combining different participatory approaches in climate action. Furthermore, AI practitioners and diverse stakeholders could encourage that algorithmic systems are deployed within new community-lead contractual models such as the Maori data sovereignty licences \cite{maori2021} and the behavioral use licensing agreements \cite{contractor2020behavioral}. For example, the Kaitiakitanga\footnote{https://papareo.nz/kaitiakitanga} affirmative action data license developed by the Maori community in New Zealand, states that any benefits derived from their data need to flow back to the source of the data. Learning from the intentionality within their data stewardship principles and process, we turn to datasets used broadly by practitioners in AI as benchmarks for assessing the utility of machine learning models. A justice-oriented approach could better empower the individuals and communities who have collected, curated, labeled, or otherwise contributed to creating datasets, have a voice in how that data is used. 

\section{Conclusion}
\label{sec6}
% Learning from the emerging themes within the field of environmental justice, here  

Building on a growing number of algorithmic harm investigations and critical scholarship, we have argued for the need to expand the assumed boundaries of algorithmic system through understanding them as ontologically indistinct from social-ecological-technological systems. Thus, our \emph{theory of change} is centered on the need and utility of SETS analysis to enable interdisciplinary collaborations in disentangling the complex nature of algorithmic impacts in \emph{a more-than human world}. Furthermore, we aim to bring awareness to potential parallels in the field of algorithmic justice and the longstanding theory and practice within the environmental justice movement. We overlay a SETS analysis framework and emerging themes in EJ to motivate a set of questions that practitioners could operationalize in the algorithmic audit process. Futhermore, we draw policy recommendations that we believe could positively contribute to considering environmental sustainability concerns within the algorithmic auditing ecosystem: (1) broadening the inputs and opening-up the outputs of an impact assessment, (2) enabling meaningful access to redress, and (3) adopting a place-based and relational approach to the process of evaluating impact. Our recommendations are rooted in lessons from the existing community of practice across algorithmic auditing, SETS analysis, and EJ. Future work will be required to link the ideas we lay out to existing policy and practical implementation in particular domains.   

\bibliographystyle{ACM-Reference-Format}
\bibliography{references}

\end{document}

%% file: tables/sets_table.tex
\begin{table}[]
\caption{Social-ecological-technological systems couplings \cite{pineda2021examining, mcphearson2021radical, markolf2018interdependent}}
\label{tab:sets-table}

\begin{tabular}{ p{2cm}p{4cm}p{4cm}p{4cm}}
 \hline
 \setlength{\tabcolsep}{25pt}
  & Social-Ecological Couplings & Ecological-Technological Couplings & Social-Technological Coupling\\ [0.5ex] 
 \hline

  \setlength{\tabcolsep}{25pt}
Key considerations
 & \emph{community-nature interactions shape social values, attitudes, behaviours, decision-making, and socio-ecological memory; in turn nature changes, reacts, and adapts}
 & \emph{extractive and polluting technological-infrastructures; responsive hybrid green-grey, circular economy, and industrial ecology approaches to restore ecological processes}
 & \emph{social norms, behaviors, values and belief systems create, maintain and replace technological-infrastructure systems; value systems shape and evolve as the interact with technological systems} 
 \\ 
 \hline

%   \setlength{\tabcolsep}{25pt}
% Implications for AI in the context of large language models 
% & \emph{..}
% & \emph{..}
% & \emph{..}
% \\ 
%  \hline
 
\end{tabular}
\label{table:1_overview}
\end{table}

%% file: tables/sets_vs_justice.tex
\begin{table}[]
\caption{Environmental justice dimensions and SETS: integrating the frameworks in the context of algorithmic auditing}
\label{tab:sets-justice-table}

\begin{tabular}{ p{2cm}p{4cm}p{4cm}p{4cm}}

 \hline

 % \setlength{\tabcolsep}{25pt}
 %   & Social-Ecological & Ecological-Technological & Social-Technological\\ [0.5ex] 
 % \hline
 % \hline

\setlength{\tabcolsep}{25pt}
\centering
EJ Dimension & \multicolumn{3}{c}{Social-ecological-technological couplings} \\
\cline{2-4}

& Social-Ecological & Ecological-Technological & Social-Technological\\ [0.5ex] 

\hline
\hline
 
\setlength{\tabcolsep}{25pt}
Practice and materiality 
% SOCIAL-ECOLOGICAL
& \emph{Does the AI system, including its potential downstream derivative applications, impact the way people interact with ecosystem services, for example, access to water, food, and energy? Does access to the AI or the resources it depends on, i.e. internet connectivity, lead to unfair outcomes? How does the AI project interfere with material resource flow and in what ways would that burden vulnerable social groups?}
% ECOLOGICAL-TECHNOLOGICAL
& \emph{What are potential impacts, risks, and threats of the material flows and ecological dependencies of AI systems and their computational infrastructure along the lifecycle - from mineral extractivism, to data centers and infrastructures, to electronic waste dumps? How do we measure and provide sufficient transparency over the carbon and ecological costs of computational infrastructure? }
% SOCIAL-TECHNOLOOGICAL
& \emph{How does the process of design, development, deployment, and auditing of an AI system interfere in the politics of sustainability of everyday life?  For example, how might data collection, labeling, and use practices interfere with the way people provide for basic human needs, i.e. access to water, food, and energy? What resources and avenues exist to meaningfully investigate or regulate the AI system and its computational infrastructure?}
\\ 
 \hline

  \setlength{\tabcolsep}{25pt}
Community, identity, and attachment 
% SOCIAL-ECOLOGICAL
& \emph{How might the AI system interfere with situated knowledges of local communities as well as their sense of collective identity and attachment to particular territories? How might the AI system interfere with displacement due to environmental disasters?}
% ECOLOGICAL-TECHNOLOGICAL
& \emph{How do material / ecological impacts of AI and computational infrastructure vary across spaces and communities? To what extent are such impacts made visible as part of the impact assessment process? What data is made visible to whom? How is data about the broader computational infrastructure made visible? }
% SOCIAL-TECHNOLOOGICAL
& \emph{What is considered data in building and auditing the AI system? Whose knowledges are taken into account? How does the AI system impact (un)intended users' individual and collective sense of identity as well as identity transitions? And what agency and power is given to engaged communities and stakeholders to shape the audit or design? }
\\ 
 \hline

  \setlength{\tabcolsep}{25pt}
Human and non-human assemblages
% SOCIAL-ECOLOGICAL
& \emph{Could we consult with environmental ethicists, environmental scientists, environmental lawyers, or grassroots organizations who could provide a critical lens for how the AI project interferes with human-nature or community-nature interactions?}
% ECOLOGICAL-TECHNOLOGICAL
& \emph{What are the risks of the AI system's computational infrastructure to land use, deforestation, biodiversity, habitat destruction, over-exploitation, and other aspects of environmental degradation?}
% SOCIAL-TECHNOLOOGICAL
& \emph{How might algorithmic bias arise outside of human populations? How might the AI system impact local social relationships to the more-than-human world within the communities where it is deployed?}
\\ 
 \hline
 
\end{tabular}
\label{table:1_overview}
\end{table}

%% file: main.bbl

\begin{thebibliography}{138}

%%% the Web address.

\ifx \showCODEN    \undefined \def \showCODEN     #1{\unskip}     \fi
\ifx \showDOI      \undefined \def \showDOI       #1{#1}\fi
\ifx \showISBNx    \undefined \def \showISBNx     #1{\unskip}     \fi
\ifx \showISBNxiii \undefined \def \showISBNxiii  #1{\unskip}     \fi
\ifx \showISSN     \undefined \def \showISSN      #1{\unskip}     \fi
\ifx \showLCCN     \undefined \def \showLCCN      #1{\unskip}     \fi
\ifx \shownote     \undefined \def \shownote      #1{#1}          \fi
\ifx \showarticletitle \undefined \def \showarticletitle #1{#1}   \fi
\ifx \showURL      \undefined \def \showURL       {\relax}        \fi
% invisible to TeX
\providecommand\bibfield[2]{#2}
\providecommand\bibinfo[2]{#2}
\providecommand\natexlab[1]{#1}
\providecommand\showeprint[2][]{arXiv:#2}

\bibitem[Agre(1997)]%
        {agre1997toward}
\bibfield{author}{\bibinfo{person}{P Agre}.} \bibinfo{year}{1997}\natexlab{}.
\newblock \showarticletitle{Toward a critical technical practice: Lessons
  learned in trying to reform AI.}
\newblock \bibinfo{journal}{\emph{G., Star, S., Turner, W., and Gasser, L.,
  eds, Social Science, Technical Systems and Cooperative Work: Beyond the Great
  Divide, Erlbaum}} (\bibinfo{year}{1997}).
\newblock


\bibitem[Agyeman et~al\mbox{.}(2003)]%
        {just2003}
\bibfield{author}{\bibinfo{person}{Julian Agyeman},
  \bibinfo{person}{Robert~Doyle Bullard}, {and} \bibinfo{person}{Bob Evans}.}
  \bibinfo{year}{2003}\natexlab{}.
\newblock \bibinfo{booktitle}{\emph{Just sustainabilities: Development in an
  unequal world}}.
\newblock \bibinfo{publisher}{MIT press}.
\newblock


\bibitem[Agyeman et~al\mbox{.}(2016)]%
        {agyeman2016trends}
\bibfield{author}{\bibinfo{person}{Julian Agyeman}, \bibinfo{person}{David
  Schlosberg}, \bibinfo{person}{Luke Craven}, {and} \bibinfo{person}{Caitlin
  Matthews}.} \bibinfo{year}{2016}\natexlab{}.
\newblock \showarticletitle{Trends and directions in environmental justice:
  from inequity to everyday life, community, and just sustainabilities}.
\newblock \bibinfo{journal}{\emph{Annual Review of Environment and Resources}}
  \bibinfo{volume}{41} (\bibinfo{year}{2016}), \bibinfo{pages}{321--340}.
\newblock


\bibitem[Ahlborg and Nightingale(2012)]%
        {ahlborg2012mismatch}
\bibfield{author}{\bibinfo{person}{Helene Ahlborg} {and}
  \bibinfo{person}{Andrea~J Nightingale}.} \bibinfo{year}{2012}\natexlab{}.
\newblock \showarticletitle{Mismatch between scales of knowledge in Nepalese
  forestry: epistemology, power, and policy implications}.
\newblock \bibinfo{journal}{\emph{Ecology and Society}} \bibinfo{volume}{17},
  \bibinfo{number}{4} (\bibinfo{year}{2012}).
\newblock


\bibitem[Ahlborg et~al\mbox{.}(2019)]%
        {ahlborg2019}
\bibfield{author}{\bibinfo{person}{Helene Ahlborg}, \bibinfo{person}{Ilse
  Ruiz-Mercado}, \bibinfo{person}{Sverker Molander}, {and}
  \bibinfo{person}{Omar Masera}.} \bibinfo{year}{2019}\natexlab{}.
\newblock \showarticletitle{Bringing technology into social-ecological systems
  research—motivations for a socio-technical-ecological systems approach}.
\newblock \bibinfo{journal}{\emph{Sustainability}} \bibinfo{volume}{11},
  \bibinfo{number}{7} (\bibinfo{year}{2019}), \bibinfo{pages}{2009}.
\newblock


\bibitem[Anderies(2014)]%
        {anderies2014embedding}
\bibfield{author}{\bibinfo{person}{John~M Anderies}.}
  \bibinfo{year}{2014}\natexlab{}.
\newblock \showarticletitle{Embedding built environments in social--ecological
  systems: resilience-based design principles}.
\newblock \bibinfo{journal}{\emph{Building Research \& Information}}
  \bibinfo{volume}{42}, \bibinfo{number}{2} (\bibinfo{year}{2014}),
  \bibinfo{pages}{130--142}.
\newblock


\bibitem[Anguelovski(2014)]%
        {anguelovski2014neighborhood}
\bibfield{author}{\bibinfo{person}{Isabelle Anguelovski}.}
  \bibinfo{year}{2014}\natexlab{}.
\newblock \bibinfo{booktitle}{\emph{Neighborhood as refuge: Community
  reconstruction, place remaking, and environmental justice in the city}}.
\newblock \bibinfo{publisher}{MIT press}.
\newblock


\bibitem[Anson et~al\mbox{.}(2022)]%
        {waterjustice}
\bibfield{author}{\bibinfo{person}{April Anson}, \bibinfo{person}{Andrea
  Ballestero}, \bibinfo{person}{Dean Chahim}, \bibinfo{person}{Theodora Dryer},
  \bibinfo{person}{Sage Gerson}, \bibinfo{person}{Matthew Henry},
  \bibinfo{person}{Hi'ilei Julia~Hobart}, \bibinfo{person}{Fushcia-Ann Hoover},
  \bibinfo{person}{J.T. Roane}, \bibinfo{person}{Amrah Salomón},
  \bibinfo{person}{Bruno Seraphin}, {and} \bibinfo{person}{Elena Sobrino}.}
  \bibinfo{year}{2022}\natexlab{}.
\newblock \showarticletitle{Water Justice and Technology: The COVID-19 Crisis,
  Computational Resource Control, and Water Relief Policy}.
\newblock \bibinfo{journal}{\emph{AI Now Institute at New York University}}
  (\bibinfo{year}{2022}).
\newblock


\bibitem[Aragon et~al\mbox{.}(2016)]%
        {aragon2016developing}
\bibfield{author}{\bibinfo{person}{Cecilia Aragon}, \bibinfo{person}{Clayton
  Hutto}, \bibinfo{person}{Andy Echenique}, \bibinfo{person}{Brittany
  Fiore-Gartland}, \bibinfo{person}{Yun Huang}, \bibinfo{person}{Jinyoung Kim},
  \bibinfo{person}{Gina Neff}, \bibinfo{person}{Wanli Xing}, {and}
  \bibinfo{person}{Joseph Bayer}.} \bibinfo{year}{2016}\natexlab{}.
\newblock \showarticletitle{Developing a research agenda for human-centered
  data science}. In \bibinfo{booktitle}{\emph{Proceedings of the 19th ACM
  Conference on Computer Supported Cooperative Work and Social Computing
  Companion}}. \bibinfo{pages}{529--535}.
\newblock


\bibitem[Balvanera et~al\mbox{.}(2017)]%
        {balvanera2017interconnected}
\bibfield{author}{\bibinfo{person}{Patricia Balvanera}, \bibinfo{person}{Rafael
  Calder{\'o}n-Contreras}, \bibinfo{person}{Antonio~J Castro},
  \bibinfo{person}{Mar{\'\i}a~R Felipe-Lucia}, \bibinfo{person}{Ilse~R
  Geijzendorffer}, \bibinfo{person}{Sander Jacobs}, \bibinfo{person}{Berta
  Martin-Lopez}, \bibinfo{person}{Ugo Arbieu}, \bibinfo{person}{Chinwe~Ifejika
  Speranza}, \bibinfo{person}{Bruno Locatelli}, {et~al\mbox{.}}}
  \bibinfo{year}{2017}\natexlab{}.
\newblock \showarticletitle{Interconnected place-based social--ecological
  research can inform global sustainability}.
\newblock \bibinfo{journal}{\emph{Current Opinion in Environmental
  Sustainability}}  \bibinfo{volume}{29} (\bibinfo{year}{2017}),
  \bibinfo{pages}{1--7}.
\newblock


\bibitem[Barca et~al\mbox{.}(2012)]%
        {barca2012case}
\bibfield{author}{\bibinfo{person}{Fabrizio Barca}, \bibinfo{person}{Philip
  McCann}, {and} \bibinfo{person}{Andr{\'e}s Rodr{\'\i}guez-Pose}.}
  \bibinfo{year}{2012}\natexlab{}.
\newblock \showarticletitle{The case for regional development intervention:
  place-based versus place-neutral approaches}.
\newblock \bibinfo{journal}{\emph{Journal of regional science}}
  \bibinfo{volume}{52}, \bibinfo{number}{1} (\bibinfo{year}{2012}),
  \bibinfo{pages}{134--152}.
\newblock


\bibitem[Bender et~al\mbox{.}(2021)]%
        {bender2021dangers}
\bibfield{author}{\bibinfo{person}{Emily~M Bender}, \bibinfo{person}{Timnit
  Gebru}, \bibinfo{person}{Angelina McMillan-Major}, {and}
  \bibinfo{person}{Shmargaret Shmitchell}.} \bibinfo{year}{2021}\natexlab{}.
\newblock \showarticletitle{On the Dangers of Stochastic Parrots: Can Language
  Models Be Too Big?}. In \bibinfo{booktitle}{\emph{Proceedings of the 2021 ACM
  conference on fairness, accountability, and transparency}}.
  \bibinfo{pages}{610--623}.
\newblock


\bibitem[Benjamin(2019)]%
        {benjamin2019assessing}
\bibfield{author}{\bibinfo{person}{Ruha Benjamin}.}
  \bibinfo{year}{2019}\natexlab{}.
\newblock \showarticletitle{Assessing risk, automating racism}.
\newblock \bibinfo{journal}{\emph{Science}} \bibinfo{volume}{366},
  \bibinfo{number}{6464} (\bibinfo{year}{2019}), \bibinfo{pages}{421--422}.
\newblock


\bibitem[Benjamin(2020)]%
        {benjamin2020race}
\bibfield{author}{\bibinfo{person}{Ruha Benjamin}.}
  \bibinfo{year}{2020}\natexlab{}.
\newblock \bibinfo{title}{Race after technology: Abolitionist tools for the new
  jim code}.
\newblock
\newblock


\bibitem[Berg(1998)]%
        {berg1998politics}
\bibfield{author}{\bibinfo{person}{Marc Berg}.}
  \bibinfo{year}{1998}\natexlab{}.
\newblock \showarticletitle{The politics of technology: On bringing social
  theory into technological design}.
\newblock \bibinfo{journal}{\emph{Science, Technology, \& Human Values}}
  \bibinfo{volume}{23}, \bibinfo{number}{4} (\bibinfo{year}{1998}),
  \bibinfo{pages}{456--490}.
\newblock


\bibitem[Berrouet et~al\mbox{.}(2018)]%
        {berrouet2018vulnerability}
\bibfield{author}{\bibinfo{person}{Lina~Mar{\'\i}a Berrouet},
  \bibinfo{person}{Jenny Machado}, {and} \bibinfo{person}{Clara
  Villegas-Palacio}.} \bibinfo{year}{2018}\natexlab{}.
\newblock \showarticletitle{Vulnerability of socio—ecological systems: A
  conceptual Framework}.
\newblock \bibinfo{journal}{\emph{Ecological Indicators}}  \bibinfo{volume}{84}
  (\bibinfo{year}{2018}), \bibinfo{pages}{632--647}.
\newblock


\bibitem[Bietti and Vatanparast(2019)]%
        {bietti2019data}
\bibfield{author}{\bibinfo{person}{Elettra Bietti} {and}
  \bibinfo{person}{Roxana Vatanparast}.} \bibinfo{year}{2019}\natexlab{}.
\newblock \showarticletitle{Data Waste}.
\newblock \bibinfo{journal}{\emph{Harvard International Law Journal}}
  (\bibinfo{year}{2019}).
\newblock


\bibitem[Biggs et~al\mbox{.}(2021)]%
        {biggs2021routledge}
\bibfield{author}{\bibinfo{person}{Reinette Biggs}, \bibinfo{person}{Alta
  De~Vos}, \bibinfo{person}{Rika Preiser}, \bibinfo{person}{Hayley Clements},
  \bibinfo{person}{Kristine Maciejewski}, {and} \bibinfo{person}{Maja
  Schl{\"u}ter}.} \bibinfo{year}{2021}\natexlab{}.
\newblock \bibinfo{booktitle}{\emph{The Routledge handbook of research methods
  for social-ecological systems}}.
\newblock \bibinfo{publisher}{Taylor \& Francis}.
\newblock


\bibitem[Bijker(1997)]%
        {bijker1997bicycles}
\bibfield{author}{\bibinfo{person}{Wiebe~E Bijker}.}
  \bibinfo{year}{1997}\natexlab{}.
\newblock \bibinfo{booktitle}{\emph{Of bicycles, bakelites, and bulbs: Toward a
  theory of sociotechnical change}}.
\newblock \bibinfo{publisher}{MIT press}.
\newblock


\bibitem[Birhane(2021)]%
        {birhane2021algorithmic}
\bibfield{author}{\bibinfo{person}{Abeba Birhane}.}
  \bibinfo{year}{2021}\natexlab{}.
\newblock \showarticletitle{Algorithmic injustice: a relational ethics
  approach}.
\newblock \bibinfo{journal}{\emph{Patterns}} \bibinfo{volume}{2},
  \bibinfo{number}{2} (\bibinfo{year}{2021}), \bibinfo{pages}{100205}.
\newblock


\bibitem[Birhane et~al\mbox{.}(2022)]%
        {birhane_values_2022}
\bibfield{author}{\bibinfo{person}{Abeba Birhane}, \bibinfo{person}{Pratyusha
  Kalluri}, \bibinfo{person}{Dallas Card}, \bibinfo{person}{William Agnew},
  \bibinfo{person}{Ravit Dotan}, {and} \bibinfo{person}{Michelle Bao}.}
  \bibinfo{year}{2022}\natexlab{}.
\newblock \showarticletitle{The {Values} {Encoded} in {Machine} {Learning}
  {Research}}. In \bibinfo{booktitle}{\emph{2022 {ACM} {Conference} on
  {Fairness}, {Accountability}, and {Transparency}}}
  \emph{(\bibinfo{series}{{FAccT} '22})}. \bibinfo{publisher}{Association for
  Computing Machinery}, \bibinfo{address}{New York, NY, USA},
  \bibinfo{pages}{173--184}.
\newblock
\showISBNx{978-1-4503-9352-2}
\urldef\tempurl%
\url{https://doi.org/10.1145/3531146.3533083}
\showDOI{\tempurl}


\bibitem[Bullard(1993)]%
        {bullard1993confronting}
\bibfield{author}{\bibinfo{person}{Robert~D Bullard}.}
  \bibinfo{year}{1993}\natexlab{}.
\newblock \bibinfo{booktitle}{\emph{Confronting environmental racism: Voices
  from the grassroots}}.
\newblock \bibinfo{publisher}{South End Press}.
\newblock


\bibitem[Buolamwini and Gebru(2018)]%
        {buolamwini2018gender}
\bibfield{author}{\bibinfo{person}{Joy Buolamwini} {and}
  \bibinfo{person}{Timnit Gebru}.} \bibinfo{year}{2018}\natexlab{}.
\newblock \showarticletitle{Gender shades: Intersectional accuracy disparities
  in commercial gender classification}. In \bibinfo{booktitle}{\emph{Conference
  on fairness, accountability and transparency}}. PMLR,
  \bibinfo{pages}{77--91}.
\newblock


\bibitem[Burgess et~al\mbox{.}(2004)]%
        {burgess2004citizens}
\bibfield{author}{\bibinfo{person}{J Burgess}, \bibinfo{person}{Jason
  Chilvers}, \bibinfo{person}{J Clark}, \bibinfo{person}{R Day},
  \bibinfo{person}{J Hunt}, \bibinfo{person}{S King}, \bibinfo{person}{P
  Simmons}, {and} \bibinfo{person}{A Stirling}.}
  \bibinfo{year}{2004}\natexlab{}.
\newblock \showarticletitle{Citizens and Specialists Deliberate Options for
  Managing the UK's Intermediate and High-Level Legacy Radioactive Waste}.
\newblock  (\bibinfo{year}{2004}).
\newblock


\bibitem[Burgess et~al\mbox{.}(2007)]%
        {burgess2007deliberative}
\bibfield{author}{\bibinfo{person}{Jacquelin Burgess}, \bibinfo{person}{Andy
  Stirling}, \bibinfo{person}{Judy Clark}, \bibinfo{person}{Gail Davies},
  \bibinfo{person}{Malcolm Eames}, \bibinfo{person}{Kristina Staley}, {and}
  \bibinfo{person}{Suzanne Williamson}.} \bibinfo{year}{2007}\natexlab{}.
\newblock \showarticletitle{Deliberative mapping: a novel analytic-deliberative
  methodology to support contested science-policy decisions}.
\newblock \bibinfo{journal}{\emph{Public Understanding of Science}}
  \bibinfo{volume}{16}, \bibinfo{number}{3} (\bibinfo{year}{2007}),
  \bibinfo{pages}{299--322}.
\newblock


\bibitem[Cagney et~al\mbox{.}(2016)]%
        {cagney2016social}
\bibfield{author}{\bibinfo{person}{Kathleen~A Cagney}, \bibinfo{person}{David
  Sterrett}, \bibinfo{person}{Jennifer Benz}, {and} \bibinfo{person}{Trevor
  Tompson}.} \bibinfo{year}{2016}\natexlab{}.
\newblock \showarticletitle{Social resources and community resilience in the
  wake of Superstorm Sandy}.
\newblock \bibinfo{journal}{\emph{PLoS One}} \bibinfo{volume}{11},
  \bibinfo{number}{8} (\bibinfo{year}{2016}), \bibinfo{pages}{e0160824}.
\newblock


\bibitem[Cath and Jansen(2021)]%
        {cath_dutch_2021}
\bibfield{author}{\bibinfo{person}{Corinne Cath} {and} \bibinfo{person}{Fieke
  Jansen}.} \bibinfo{year}{2021}\natexlab{}.
\newblock \bibinfo{title}{Dutch {Comfort}: {The} limits of {AI} governance
  through municipal registers}.
\newblock
\newblock
\urldef\tempurl%
\url{https://doi.org/10.48550/arXiv.2109.02944}
\showDOI{\tempurl}
\newblock
\shownote{arXiv:2109.02944 [cs]}.


\bibitem[Chancellor et~al\mbox{.}(2019)]%
        {chancellor2019human}
\bibfield{author}{\bibinfo{person}{Stevie Chancellor}, \bibinfo{person}{Eric~PS
  Baumer}, {and} \bibinfo{person}{Munmun De~Choudhury}.}
  \bibinfo{year}{2019}\natexlab{}.
\newblock \showarticletitle{Who is the" human" in human-centered machine
  learning: The case of predicting mental health from social media}.
\newblock \bibinfo{journal}{\emph{Proceedings of the ACM on Human-Computer
  Interaction}} \bibinfo{volume}{3}, \bibinfo{number}{CSCW}
  (\bibinfo{year}{2019}), \bibinfo{pages}{1--32}.
\newblock


\bibitem[Clark and Harley(2020)]%
        {clark2020sustainability}
\bibfield{author}{\bibinfo{person}{William~C Clark} {and}
  \bibinfo{person}{Alicia~G Harley}.} \bibinfo{year}{2020}\natexlab{}.
\newblock \showarticletitle{Sustainability science: Toward a synthesis}.
\newblock \bibinfo{journal}{\emph{Annual Review of Environment and Resources}}
  \bibinfo{volume}{45} (\bibinfo{year}{2020}), \bibinfo{pages}{331--386}.
\newblock


\bibitem[Clutton-Brock et~al\mbox{.}(2021)]%
        {clutton2021climate}
\bibfield{author}{\bibinfo{person}{Peter Clutton-Brock}, \bibinfo{person}{David
  Rolnick}, \bibinfo{person}{Priya~L Donti}, {and} \bibinfo{person}{Lynn
  Kaack}.} \bibinfo{year}{2021}\natexlab{}.
\newblock \bibinfo{booktitle}{\emph{Climate Change and AI. Recommendations for
  Government Action}}.
\newblock \bibinfo{type}{{T}echnical {R}eport}. \bibinfo{institution}{GPAI,
  Climate Change AI, Centre for AI \& Climate}.
\newblock


\bibitem[Coeckelbergh(2021)]%
        {coeckelbergh2021ai}
\bibfield{author}{\bibinfo{person}{Mark Coeckelbergh}.}
  \bibinfo{year}{2021}\natexlab{}.
\newblock \showarticletitle{AI for climate: freedom, justice, and other ethical
  and political challenges}.
\newblock \bibinfo{journal}{\emph{AI and Ethics}} \bibinfo{volume}{1},
  \bibinfo{number}{1} (\bibinfo{year}{2021}), \bibinfo{pages}{67--72}.
\newblock


\bibitem[Contractor et~al\mbox{.}(2020)]%
        {contractor2020behavioral}
\bibfield{author}{\bibinfo{person}{Danish Contractor}, \bibinfo{person}{Daniel
  McDuff}, \bibinfo{person}{Julia Haines}, \bibinfo{person}{Jenny Lee},
  \bibinfo{person}{Christopher Hines}, {and} \bibinfo{person}{Brent Hecht}.}
  \bibinfo{year}{2020}\natexlab{}.
\newblock \showarticletitle{Behavioral use licensing for responsible AI}.
\newblock \bibinfo{journal}{\emph{arXiv preprint arXiv:2011.03116}}
  (\bibinfo{year}{2020}).
\newblock


\bibitem[Cooper et~al\mbox{.}(2022)]%
        {cooper2022accountability}
\bibfield{author}{\bibinfo{person}{A~Feder Cooper}, \bibinfo{person}{Emanuel
  Moss}, \bibinfo{person}{Benjamin Laufer}, {and} \bibinfo{person}{Helen
  Nissenbaum}.} \bibinfo{year}{2022}\natexlab{}.
\newblock \showarticletitle{Accountability in an algorithmic society:
  relationality, responsibility, and robustness in machine learning}. In
  \bibinfo{booktitle}{\emph{2022 ACM Conference on Fairness, Accountability,
  and Transparency}}. \bibinfo{pages}{864--876}.
\newblock


\bibitem[Costanza-Chock(2020)]%
        {costanza2020design}
\bibfield{author}{\bibinfo{person}{Sasha Costanza-Chock}.}
  \bibinfo{year}{2020}\natexlab{}.
\newblock \bibinfo{booktitle}{\emph{Design justice: Community-led practices to
  build the worlds we need}}.
\newblock \bibinfo{publisher}{The MIT Press}.
\newblock


\bibitem[Costanza-Chock et~al\mbox{.}(2022)]%
        {costanza2022audits}
\bibfield{author}{\bibinfo{person}{Sasha Costanza-Chock},
  \bibinfo{person}{Inioluwa~Deborah Raji}, {and} \bibinfo{person}{Joy
  Buolamwini}.} \bibinfo{year}{2022}\natexlab{}.
\newblock \showarticletitle{Who Audits the Auditors? Recommendations from a
  field scan of the algorithmic auditing ecosystem}. In
  \bibinfo{booktitle}{\emph{2022 ACM Conference on Fairness, Accountability,
  and Transparency}}. \bibinfo{pages}{1571--1583}.
\newblock


\bibitem[Coulombel et~al\mbox{.}(2019)]%
        {coulombel2019substantial}
\bibfield{author}{\bibinfo{person}{Nicolas Coulombel},
  \bibinfo{person}{Virginie Boutueil}, \bibinfo{person}{Liu Liu},
  \bibinfo{person}{Vincent Viguie}, {and} \bibinfo{person}{Biao Yin}.}
  \bibinfo{year}{2019}\natexlab{}.
\newblock \showarticletitle{Substantial rebound effects in urban ridesharing:
  Simulating travel decisions in Paris, France}.
\newblock \bibinfo{journal}{\emph{Transportation Research Part D: Transport and
  Environment}}  \bibinfo{volume}{71} (\bibinfo{year}{2019}),
  \bibinfo{pages}{110--126}.
\newblock


\bibitem[Crawford and Joler(2018)]%
        {crawford2018anatomy}
\bibfield{author}{\bibinfo{person}{Kate Crawford} {and} \bibinfo{person}{Vladan
  Joler}.} \bibinfo{year}{2018}\natexlab{}.
\newblock \showarticletitle{Anatomy of an AI System}.
\newblock \bibinfo{journal}{\emph{Retrieved September}}  \bibinfo{volume}{18}
  (\bibinfo{year}{2018}), \bibinfo{pages}{2018}.
\newblock


\bibitem[Cutter(2002)]%
        {cutter2002american}
\bibfield{author}{\bibinfo{person}{Susan~L Cutter}.}
  \bibinfo{year}{2002}\natexlab{}.
\newblock \bibinfo{booktitle}{\emph{American hazardscapes: The regionalization
  of hazards and disasters}}.
\newblock \bibinfo{publisher}{Joseph Henry Press}.
\newblock


\bibitem[Davis et~al\mbox{.}(2021)]%
        {davis2021algorithmic}
\bibfield{author}{\bibinfo{person}{Jenny~L Davis}, \bibinfo{person}{Apryl
  Williams}, {and} \bibinfo{person}{Michael~W Yang}.}
  \bibinfo{year}{2021}\natexlab{}.
\newblock \showarticletitle{Algorithmic reparation}.
\newblock \bibinfo{journal}{\emph{Big Data \& Society}} \bibinfo{volume}{8},
  \bibinfo{number}{2} (\bibinfo{year}{2021}),
  \bibinfo{pages}{20539517211044808}.
\newblock


\bibitem[Dobbe et~al\mbox{.}(2021)]%
        {dobbe2021hard}
\bibfield{author}{\bibinfo{person}{Roel Dobbe}, \bibinfo{person}{Thomas~Krendl
  Gilbert}, {and} \bibinfo{person}{Yonatan Mintz}.}
  \bibinfo{year}{2021}\natexlab{}.
\newblock \showarticletitle{Hard choices in artificial intelligence}.
\newblock \bibinfo{journal}{\emph{Artificial Intelligence}}
  \bibinfo{volume}{300} (\bibinfo{year}{2021}), \bibinfo{pages}{103555}.
\newblock


\bibitem[Dobbe and Whittaker(2019)]%
        {dobbe2019ai}
\bibfield{author}{\bibinfo{person}{Roel Dobbe} {and} \bibinfo{person}{Meredith
  Whittaker}.} \bibinfo{year}{2019}\natexlab{}.
\newblock \showarticletitle{AI and Climate Change: How they’re connected, and
  what we can do about it}.
\newblock \bibinfo{journal}{\emph{AI Now Institute}}  \bibinfo{volume}{17}
  (\bibinfo{year}{2019}).
\newblock


\bibitem[Dryer(2020)]%
        {dryer_ai_2020}
\bibfield{author}{\bibinfo{person}{Theodora Dryer}.}
  \bibinfo{year}{2020}\natexlab{}.
\newblock \showarticletitle{{AI} and its {Computing} {Landscapes}: {Water}
  {Data}, {Climate} {Control}, and {Agricultural} {Technology}}.
\newblock \bibinfo{journal}{\emph{Bulletin of the American Physical Society}}
  \bibinfo{volume}{65} (\bibinfo{year}{2020}).
\newblock
\newblock
\shownote{Publisher: APS}.


\bibitem[Ehsan and Riedl(2020)]%
        {ehsan2020human}
\bibfield{author}{\bibinfo{person}{Upol Ehsan} {and} \bibinfo{person}{Mark~O
  Riedl}.} \bibinfo{year}{2020}\natexlab{}.
\newblock \showarticletitle{Human-centered explainable ai: towards a reflective
  sociotechnical approach}. In \bibinfo{booktitle}{\emph{International
  Conference on Human-Computer Interaction}}. Springer,
  \bibinfo{pages}{449--466}.
\newblock


\bibitem[Eubanks(2018)]%
        {eubanks2018automating}
\bibfield{author}{\bibinfo{person}{Virginia Eubanks}.}
  \bibinfo{year}{2018}\natexlab{}.
\newblock \bibinfo{booktitle}{\emph{Automating inequality: How high-tech tools
  profile, police, and punish the poor}}.
\newblock \bibinfo{publisher}{St. Martin's Press}.
\newblock


\bibitem[Foucault(1991)]%
        {foucault1991foucault}
\bibfield{author}{\bibinfo{person}{Michel Foucault}.}
  \bibinfo{year}{1991}\natexlab{}.
\newblock \bibinfo{booktitle}{\emph{The Foucault effect: Studies in
  governmentality}}.
\newblock \bibinfo{publisher}{University of Chicago Press}.
\newblock


\bibitem[Foxon et~al\mbox{.}(2009)]%
        {foxon2009governing}
\bibfield{author}{\bibinfo{person}{Timothy~J Foxon}, \bibinfo{person}{Mark~S
  Reed}, {and} \bibinfo{person}{Lindsay~C Stringer}.}
  \bibinfo{year}{2009}\natexlab{}.
\newblock \showarticletitle{Governing long-term social--ecological change: what
  can the adaptive management and transition management approaches learn from
  each other?}
\newblock \bibinfo{journal}{\emph{Environmental Policy and Governance}}
  \bibinfo{volume}{19}, \bibinfo{number}{1} (\bibinfo{year}{2009}),
  \bibinfo{pages}{3--20}.
\newblock


\bibitem[Ganesh and Moss(2022)]%
        {ganesh2022resistance}
\bibfield{author}{\bibinfo{person}{Maya~Indira Ganesh} {and}
  \bibinfo{person}{Emanuel Moss}.} \bibinfo{year}{2022}\natexlab{}.
\newblock \showarticletitle{Resistance and refusal to algorithmic harms:
  Varieties of ‘knowledge projects’}.
\newblock \bibinfo{journal}{\emph{Media International Australia}}
  \bibinfo{volume}{183}, \bibinfo{number}{1} (\bibinfo{year}{2022}),
  \bibinfo{pages}{90--106}.
\newblock


\bibitem[Gansky and McDonald(2022)]%
        {gansky_counterfacctual_2022}
\bibfield{author}{\bibinfo{person}{Ben Gansky} {and} \bibinfo{person}{Sean
  McDonald}.} \bibinfo{year}{2022}\natexlab{}.
\newblock \showarticletitle{{CounterFAccTual}: {How} {FAccT} {Undermines} {Its}
  {Organizing} {Principles}}. In \bibinfo{booktitle}{\emph{2022 {ACM}
  {Conference} on {Fairness}, {Accountability}, and {Transparency}}}
  \emph{(\bibinfo{series}{{FAccT} '22})}. \bibinfo{publisher}{Association for
  Computing Machinery}, \bibinfo{address}{New York, NY, USA},
  \bibinfo{pages}{1982--1992}.
\newblock
\showISBNx{978-1-4503-9352-2}
\urldef\tempurl%
\url{https://doi.org/10.1145/3531146.3533241}
\showDOI{\tempurl}


\bibitem[Goldstein et~al\mbox{.}(2015)]%
        {goldstein2015narrating}
\bibfield{author}{\bibinfo{person}{Bruce~Evan Goldstein},
  \bibinfo{person}{Anne~Taufen Wessells}, \bibinfo{person}{Raul Lejano}, {and}
  \bibinfo{person}{William Butler}.} \bibinfo{year}{2015}\natexlab{}.
\newblock \showarticletitle{Narrating resilience: Transforming urban systems
  through collaborative storytelling}.
\newblock \bibinfo{journal}{\emph{Urban Studies}} \bibinfo{volume}{52},
  \bibinfo{number}{7} (\bibinfo{year}{2015}), \bibinfo{pages}{1285--1303}.
\newblock


\bibitem[Green and Viljoen(2020)]%
        {green2020algorithmic}
\bibfield{author}{\bibinfo{person}{Ben Green} {and} \bibinfo{person}{Salome
  Viljoen}.} \bibinfo{year}{2020}\natexlab{}.
\newblock \showarticletitle{Algorithmic realism: expanding the boundaries of
  algorithmic thought}. In \bibinfo{booktitle}{\emph{Proceedings of the 2020
  conference on fairness, accountability, and transparency}}.
  \bibinfo{pages}{19--31}.
\newblock


\bibitem[Haimson et~al\mbox{.}(2015)]%
        {haimson2015online}
\bibfield{author}{\bibinfo{person}{Oliver~L Haimson}, \bibinfo{person}{Anne~E
  Bowser}, \bibinfo{person}{Edward~F Melcer}, {and}
  \bibinfo{person}{Elizabeth~F Churchill}.} \bibinfo{year}{2015}\natexlab{}.
\newblock \showarticletitle{Online inspiration and exploration for identity
  reinvention}. In \bibinfo{booktitle}{\emph{Proceedings of the 33rd annual ACM
  conference on human factors in computing systems}}.
  \bibinfo{pages}{3809--3818}.
\newblock


\bibitem[Haimson and Hoffmann(2016)]%
        {haimson2016constructing}
\bibfield{author}{\bibinfo{person}{Oliver~L Haimson} {and}
  \bibinfo{person}{Anna~Lauren Hoffmann}.} \bibinfo{year}{2016}\natexlab{}.
\newblock \showarticletitle{Constructing and enforcing" authentic" identity
  online: Facebook, real names, and non-normative identities}.
\newblock \bibinfo{journal}{\emph{First Monday}} (\bibinfo{year}{2016}).
\newblock


\bibitem[Hanna et~al\mbox{.}(2020)]%
        {hanna2020towards}
\bibfield{author}{\bibinfo{person}{Alex Hanna}, \bibinfo{person}{Emily Denton},
  \bibinfo{person}{Andrew Smart}, {and} \bibinfo{person}{Jamila Smith-Loud}.}
  \bibinfo{year}{2020}\natexlab{}.
\newblock \showarticletitle{Towards a critical race methodology in algorithmic
  fairness}. In \bibinfo{booktitle}{\emph{Proceedings of the 2020 conference on
  fairness, accountability, and transparency}}. \bibinfo{pages}{501--512}.
\newblock


\bibitem[Haraway(2006)]%
        {haraway2006cyborg}
\bibfield{author}{\bibinfo{person}{Donna Haraway}.}
  \bibinfo{year}{2006}\natexlab{}.
\newblock \showarticletitle{A cyborg manifesto: Science, technology, and
  socialist-feminism in the late 20th century}.
\newblock In \bibinfo{booktitle}{\emph{The international handbook of virtual
  learning environments}}. \bibinfo{publisher}{Springer},
  \bibinfo{pages}{117--158}.
\newblock


\bibitem[Hasinoff and Schneider(2022)]%
        {hasinoff2022scalability}
\bibfield{author}{\bibinfo{person}{Amy~A Hasinoff} {and}
  \bibinfo{person}{Nathan Schneider}.} \bibinfo{year}{2022}\natexlab{}.
\newblock \showarticletitle{From Scalability to Subsidiarity in Addressing
  Online Harm}.
\newblock \bibinfo{journal}{\emph{Social Media+ Society}} \bibinfo{volume}{8},
  \bibinfo{number}{3} (\bibinfo{year}{2022}),
  \bibinfo{pages}{20563051221126041}.
\newblock


\bibitem[Hecht et~al\mbox{.}(2021)]%
        {hecht2021s}
\bibfield{author}{\bibinfo{person}{Brent Hecht}, \bibinfo{person}{Lauren
  Wilcox}, \bibinfo{person}{Jeffrey~P Bigham}, \bibinfo{person}{Johannes
  Sch{\"o}ning}, \bibinfo{person}{Ehsan Hoque}, \bibinfo{person}{Jason Ernst},
  \bibinfo{person}{Yonatan Bisk}, \bibinfo{person}{Luigi De~Russis},
  \bibinfo{person}{Lana Yarosh}, \bibinfo{person}{Bushra Anjum},
  {et~al\mbox{.}}} \bibinfo{year}{2021}\natexlab{}.
\newblock \showarticletitle{It's time to do something: Mitigating the negative
  impacts of computing through a change to the peer review process}.
\newblock \bibinfo{journal}{\emph{arXiv preprint arXiv:2112.09544}}
  (\bibinfo{year}{2021}).
\newblock


\bibitem[HG~Solomon and Baio(2020)]%
        {hg2020argument}
\bibfield{author}{\bibinfo{person}{Lucy HG~Solomon} {and}
  \bibinfo{person}{Cesar Baio}.} \bibinfo{year}{2020}\natexlab{}.
\newblock \showarticletitle{An Argument for an Ecosystemic AI: Articulating
  Connections across Prehuman and Posthuman Intelligences}.
\newblock \bibinfo{journal}{\emph{International Journal of Community
  Well-Being}} \bibinfo{volume}{3}, \bibinfo{number}{4} (\bibinfo{year}{2020}),
  \bibinfo{pages}{559--584}.
\newblock


\bibitem[Hoffmann(2019)]%
        {hoffmann2019fairness}
\bibfield{author}{\bibinfo{person}{Anna~Lauren Hoffmann}.}
  \bibinfo{year}{2019}\natexlab{}.
\newblock \showarticletitle{Where fairness fails: data, algorithms, and the
  limits of antidiscrimination discourse}.
\newblock \bibinfo{journal}{\emph{Information, Communication \& Society}}
  \bibinfo{volume}{22}, \bibinfo{number}{7} (\bibinfo{year}{2019}),
  \bibinfo{pages}{900--915}.
\newblock


\bibitem[Holifield(2001)]%
        {holifield2001defining}
\bibfield{author}{\bibinfo{person}{Ryan Holifield}.}
  \bibinfo{year}{2001}\natexlab{}.
\newblock \showarticletitle{Defining environmental justice and environmental
  racism}.
\newblock \bibinfo{journal}{\emph{Urban geography}} \bibinfo{volume}{22},
  \bibinfo{number}{1} (\bibinfo{year}{2001}), \bibinfo{pages}{78--90}.
\newblock


\bibitem[Itten and Mouter(2022a)]%
        {su14084656}
\bibfield{author}{\bibinfo{person}{Anatol Itten} {and} \bibinfo{person}{Niek
  Mouter}.} \bibinfo{year}{2022}\natexlab{a}.
\newblock \showarticletitle{When Digital Mass Participation Meets Citizen
  Deliberation: Combining Mini- and Maxi-Publics in Climate Policy-Making}.
\newblock \bibinfo{journal}{\emph{Sustainability}} \bibinfo{volume}{14},
  \bibinfo{number}{8} (\bibinfo{year}{2022}).
\newblock
\showISSN{2071-1050}
\urldef\tempurl%
\url{https://www.mdpi.com/2071-1050/14/8/4656}
\showURL{%
\tempurl}


\bibitem[Itten and Mouter(2022b)]%
        {itten2022digital}
\bibfield{author}{\bibinfo{person}{Anatol Itten} {and} \bibinfo{person}{Niek
  Mouter}.} \bibinfo{year}{2022}\natexlab{b}.
\newblock \showarticletitle{When digital mass participation meets citizen
  deliberation: combining mini-and maxi-publics in climate policy-making}.
\newblock \bibinfo{journal}{\emph{Sustainability}} \bibinfo{volume}{14},
  \bibinfo{number}{8} (\bibinfo{year}{2022}), \bibinfo{pages}{4656}.
\newblock


\bibitem[Jasanoff(2015)]%
        {jasanoff2015future}
\bibfield{author}{\bibinfo{person}{Sheila Jasanoff}.}
  \bibinfo{year}{2015}\natexlab{}.
\newblock \showarticletitle{Future imperfect: Science, technology, and the
  imaginations of modernity}.
\newblock \bibinfo{journal}{\emph{Dreamscapes of modernity: Sociotechnical
  imaginaries and the fabrication of power}} (\bibinfo{year}{2015}),
  \bibinfo{pages}{1--33}.
\newblock


\bibitem[Kaack et~al\mbox{.}(2022)]%
        {kaack2022aligning}
\bibfield{author}{\bibinfo{person}{Lynn~H Kaack}, \bibinfo{person}{Priya~L
  Donti}, \bibinfo{person}{Emma Strubell}, \bibinfo{person}{George Kamiya},
  \bibinfo{person}{Felix Creutzig}, {and} \bibinfo{person}{David Rolnick}.}
  \bibinfo{year}{2022}\natexlab{}.
\newblock \showarticletitle{Aligning artificial intelligence with climate
  change mitigation}.
\newblock \bibinfo{journal}{\emph{Nature Climate Change}} \bibinfo{volume}{12},
  \bibinfo{number}{6} (\bibinfo{year}{2022}), \bibinfo{pages}{518--527}.
\newblock


\bibitem[Katell et~al\mbox{.}(2020)]%
        {katell2020toward}
\bibfield{author}{\bibinfo{person}{Michael Katell}, \bibinfo{person}{Meg
  Young}, \bibinfo{person}{Dharma Dailey}, \bibinfo{person}{Bernease Herman},
  \bibinfo{person}{Vivian Guetler}, \bibinfo{person}{Aaron Tam},
  \bibinfo{person}{Corinne Bintz}, \bibinfo{person}{Daniella Raz}, {and}
  \bibinfo{person}{PM Krafft}.} \bibinfo{year}{2020}\natexlab{}.
\newblock \showarticletitle{Toward situated interventions for algorithmic
  equity: lessons from the field}. In \bibinfo{booktitle}{\emph{Proceedings of
  the 2020 conference on fairness, accountability, and transparency}}.
  \bibinfo{pages}{45--55}.
\newblock


\bibitem[Keyes et~al\mbox{.}(2021)]%
        {keyes2021truth}
\bibfield{author}{\bibinfo{person}{Os Keyes}, \bibinfo{person}{Zo{\"e} Hitzig},
  {and} \bibinfo{person}{Mwenza Blell}.} \bibinfo{year}{2021}\natexlab{}.
\newblock \showarticletitle{Truth from the machine: artificial intelligence and
  the materialization of identity}.
\newblock \bibinfo{journal}{\emph{Interdisciplinary Science Reviews}}
  \bibinfo{volume}{46}, \bibinfo{number}{1-2} (\bibinfo{year}{2021}),
  \bibinfo{pages}{158--175}.
\newblock


\bibitem[Kuglitsch et~al\mbox{.}(2022)]%
        {kuglitsch2022artificial}
\bibfield{author}{\bibinfo{person}{Monique Kuglitsch}, \bibinfo{person}{Arif
  Albayrak}, \bibinfo{person}{Ra{\'u}l Aquino}, \bibinfo{person}{Allison
  Craddock}, \bibinfo{person}{Jaselle Edward-Gill}, \bibinfo{person}{Rinku
  Kanwar}, \bibinfo{person}{Anirudh Koul}, \bibinfo{person}{Jackie Ma},
  \bibinfo{person}{Alejandro Marti}, \bibinfo{person}{Mythili Menon},
  {et~al\mbox{.}}} \bibinfo{year}{2022}\natexlab{}.
\newblock \showarticletitle{Artificial intelligence for disaster risk
  reduction: opportunities, challenges, and prospects}.
\newblock \bibinfo{journal}{\emph{World Meteorological Organization}}
  \bibinfo{volume}{71}, \bibinfo{number}{1} (\bibinfo{year}{2022}).
\newblock


\bibitem[Kukutai and Taylor(2016)]%
        {kukutai2016indigenous}
\bibfield{author}{\bibinfo{person}{Tahu Kukutai} {and} \bibinfo{person}{John
  Taylor}.} \bibinfo{year}{2016}\natexlab{}.
\newblock \bibinfo{booktitle}{\emph{Indigenous data sovereignty: Toward an
  agenda}}.
\newblock \bibinfo{publisher}{ANU press}.
\newblock


\bibitem[Leach et~al\mbox{.}(2010)]%
        {leach2010dynamic}
\bibfield{author}{\bibinfo{person}{Melissa Leach},
  \bibinfo{person}{Andrew~Charles Stirling}, {and} \bibinfo{person}{Ian
  Scoones}.} \bibinfo{year}{2010}\natexlab{}.
\newblock \bibinfo{booktitle}{\emph{Dynamic Sustainabilities: Technology,
  Environment, Social Justice}}.
\newblock \bibinfo{publisher}{Routledge}.
\newblock


\bibitem[Lejano(2019)]%
        {lejano2019relationality}
\bibfield{author}{\bibinfo{person}{Raul~P Lejano}.}
  \bibinfo{year}{2019}\natexlab{}.
\newblock \showarticletitle{Relationality and social--ecological systems: Going
  beyond or behind sustainability and resilience}.
\newblock \bibinfo{journal}{\emph{Sustainability}} \bibinfo{volume}{11},
  \bibinfo{number}{10} (\bibinfo{year}{2019}), \bibinfo{pages}{2760}.
\newblock


\bibitem[Lewis et~al\mbox{.}(2018)]%
        {lewis2018making}
\bibfield{author}{\bibinfo{person}{Jason~Edward Lewis},
  \bibinfo{person}{Noelani Arista}, \bibinfo{person}{Archer Pechawis}, {and}
  \bibinfo{person}{Suzanne Kite}.} \bibinfo{year}{2018}\natexlab{}.
\newblock \showarticletitle{Making kin with the machines}.
\newblock \bibinfo{journal}{\emph{Journal of Design and Science}}
  (\bibinfo{year}{2018}).
\newblock


\bibitem[Lottick et~al\mbox{.}(2019)]%
        {lottick2019energy}
\bibfield{author}{\bibinfo{person}{Kadan Lottick}, \bibinfo{person}{Silvia
  Susai}, \bibinfo{person}{Sorelle~A Friedler}, {and}
  \bibinfo{person}{Jonathan~P Wilson}.} \bibinfo{year}{2019}\natexlab{}.
\newblock \showarticletitle{Energy Usage Reports: Environmental awareness as
  part of algorithmic accountability}.
\newblock \bibinfo{journal}{\emph{arXiv preprint arXiv:1911.08354}}
  (\bibinfo{year}{2019}).
\newblock


\bibitem[Luccioni et~al\mbox{.}(2022)]%
        {luccioni2022estimating}
\bibfield{author}{\bibinfo{person}{Alexandra~Sasha Luccioni},
  \bibinfo{person}{Sylvain Viguier}, {and} \bibinfo{person}{Anne-Laure
  Ligozat}.} \bibinfo{year}{2022}\natexlab{}.
\newblock \showarticletitle{Estimating the Carbon Footprint of BLOOM, a 176B
  Parameter Language Model}.
\newblock \bibinfo{journal}{\emph{arXiv preprint arXiv:2211.02001}}
  (\bibinfo{year}{2022}).
\newblock


\bibitem[Malik and Malik(2021)]%
        {malik2021critical}
\bibfield{author}{\bibinfo{person}{Maya Malik} {and} \bibinfo{person}{Momin~M
  Malik}.} \bibinfo{year}{2021}\natexlab{}.
\newblock \showarticletitle{Critical technical awakenings}.
\newblock \bibinfo{journal}{\emph{Journal of Social Computing}}
  \bibinfo{volume}{2}, \bibinfo{number}{4} (\bibinfo{year}{2021}),
  \bibinfo{pages}{365--384}.
\newblock


\bibitem[Markolf et~al\mbox{.}(2018)]%
        {markolf2018interdependent}
\bibfield{author}{\bibinfo{person}{Samuel~A Markolf},
  \bibinfo{person}{Mikhail~V Chester}, \bibinfo{person}{Daniel~A Eisenberg},
  \bibinfo{person}{David~M Iwaniec}, \bibinfo{person}{Cliff~I Davidson},
  \bibinfo{person}{Rae Zimmerman}, \bibinfo{person}{Thaddeus~R Miller},
  \bibinfo{person}{Benjamin~L Ruddell}, {and} \bibinfo{person}{Heejun Chang}.}
  \bibinfo{year}{2018}\natexlab{}.
\newblock \showarticletitle{Interdependent infrastructure as linked social,
  ecological, and technological systems (SETSs) to address lock-in and enhance
  resilience}.
\newblock \bibinfo{journal}{\emph{Earth's Future}} \bibinfo{volume}{6},
  \bibinfo{number}{12} (\bibinfo{year}{2018}), \bibinfo{pages}{1638--1659}.
\newblock


\bibitem[Matus and Veale(2022)]%
        {matus2022certification}
\bibfield{author}{\bibinfo{person}{Kira~JM Matus} {and}
  \bibinfo{person}{Michael Veale}.} \bibinfo{year}{2022}\natexlab{}.
\newblock \showarticletitle{Certification systems for machine learning: Lessons
  from sustainability}.
\newblock \bibinfo{journal}{\emph{Regulation \& Governance}}
  \bibinfo{volume}{16}, \bibinfo{number}{1} (\bibinfo{year}{2022}),
  \bibinfo{pages}{177--196}.
\newblock


\bibitem[McGinnis and Ostrom(2014)]%
        {mcginnis2014social}
\bibfield{author}{\bibinfo{person}{Michael~D McGinnis} {and}
  \bibinfo{person}{Elinor Ostrom}.} \bibinfo{year}{2014}\natexlab{}.
\newblock \showarticletitle{Social-ecological system framework: initial changes
  and continuing challenges}.
\newblock \bibinfo{journal}{\emph{Ecology and society}} \bibinfo{volume}{19},
  \bibinfo{number}{2} (\bibinfo{year}{2014}).
\newblock


\bibitem[McGovern et~al\mbox{.}(2021)]%
        {mcgovern2021need}
\bibfield{author}{\bibinfo{person}{Amy McGovern}, \bibinfo{person}{Imme
  Ebert-Uphoff}, \bibinfo{person}{David~John Gagne~II}, {and}
  \bibinfo{person}{Ann Bostrom}.} \bibinfo{year}{2021}\natexlab{}.
\newblock \showarticletitle{The Need for Ethical, Responsible, and Trustworthy
  Artificial Intelligence for Environmental Sciences}.
\newblock \bibinfo{journal}{\emph{arXiv preprint arXiv:2112.08453}}
  (\bibinfo{year}{2021}).
\newblock


\bibitem[McPhearson et~al\mbox{.}(2021)]%
        {mcphearson2021radical}
\bibfield{author}{\bibinfo{person}{Timon McPhearson},
  \bibinfo{person}{Christopher M~Raymond}, \bibinfo{person}{Natalie Gulsrud},
  \bibinfo{person}{Christian Albert}, \bibinfo{person}{Neil Coles},
  \bibinfo{person}{Nora Fagerholm}, \bibinfo{person}{Michiru Nagatsu},
  \bibinfo{person}{Anton~Stahl Olafsson}, \bibinfo{person}{Niko Soininen},
  {and} \bibinfo{person}{Kati Vierikko}.} \bibinfo{year}{2021}\natexlab{}.
\newblock \showarticletitle{Radical changes are needed for transformations to a
  good Anthropocene}.
\newblock \bibinfo{journal}{\emph{Npj Urban Sustainability}}
  \bibinfo{volume}{1}, \bibinfo{number}{1} (\bibinfo{year}{2021}),
  \bibinfo{pages}{1--13}.
\newblock


\bibitem[McPhearson et~al\mbox{.}(2016)]%
        {mcphearson2016advancing}
\bibfield{author}{\bibinfo{person}{Timon McPhearson},
  \bibinfo{person}{Steward~TA Pickett}, \bibinfo{person}{Nancy~B Grimm},
  \bibinfo{person}{Jari Niemel{\"a}}, \bibinfo{person}{Marina Alberti},
  \bibinfo{person}{Thomas Elmqvist}, \bibinfo{person}{Christiane Weber},
  \bibinfo{person}{Dagmar Haase}, \bibinfo{person}{J{\"u}rgen Breuste}, {and}
  \bibinfo{person}{Salman Qureshi}.} \bibinfo{year}{2016}\natexlab{}.
\newblock \showarticletitle{Advancing urban ecology toward a science of
  cities}.
\newblock \bibinfo{journal}{\emph{BioScience}} \bibinfo{volume}{66},
  \bibinfo{number}{3} (\bibinfo{year}{2016}), \bibinfo{pages}{198--212}.
\newblock


\bibitem[Menton et~al\mbox{.}(2020)]%
        {menton2020environmental}
\bibfield{author}{\bibinfo{person}{Mary Menton}, \bibinfo{person}{Carlos
  Larrea}, \bibinfo{person}{Sara Latorre}, \bibinfo{person}{Joan
  Martinez-Alier}, \bibinfo{person}{Mika Peck}, \bibinfo{person}{Leah Temper},
  {and} \bibinfo{person}{Mariana Walter}.} \bibinfo{year}{2020}\natexlab{}.
\newblock \showarticletitle{Environmental justice and the SDGs: from synergies
  to gaps and contradictions}.
\newblock \bibinfo{journal}{\emph{Sustainability Science}}
  \bibinfo{volume}{15}, \bibinfo{number}{6} (\bibinfo{year}{2020}),
  \bibinfo{pages}{1621--1636}.
\newblock


\bibitem[Metcalf et~al\mbox{.}(2021)]%
        {jake2021}
\bibfield{author}{\bibinfo{person}{Jacob Metcalf}, \bibinfo{person}{Emanuel
  Moss}, \bibinfo{person}{Elizabeth~Anne Watkins}, \bibinfo{person}{Ranjit
  Singh}, {and} \bibinfo{person}{Madeleine~Clare Elish}.}
  \bibinfo{year}{2021}\natexlab{}.
\newblock \showarticletitle{Algorithmic impact assessments and accountability:
  The co-construction of impacts}. In \bibinfo{booktitle}{\emph{Proceedings of
  the 2021 ACM Conference on Fairness, Accountability, and Transparency}}.
  \bibinfo{pages}{735--746}.
\newblock


\bibitem[Mhlambi(2020)]%
        {mhlambi2020rationality}
\bibfield{author}{\bibinfo{person}{Sabelo Mhlambi}.}
  \bibinfo{year}{2020}\natexlab{}.
\newblock \showarticletitle{From rationality to relationality: ubuntu as an
  ethical and human rights framework for artificial intelligence governance}.
\newblock \bibinfo{journal}{\emph{Carr Center for Human Rights Policy
  Discussion Paper Series}}  \bibinfo{volume}{9} (\bibinfo{year}{2020}).
\newblock


\bibitem[Mohai et~al\mbox{.}(2009)]%
        {mohai2009environmental}
\bibfield{author}{\bibinfo{person}{Paul Mohai}, \bibinfo{person}{David Pellow},
  {and} \bibinfo{person}{J~Timmons Roberts}.} \bibinfo{year}{2009}\natexlab{}.
\newblock \showarticletitle{Environmental justice}.
\newblock \bibinfo{journal}{\emph{Annual review of environment and resources}}
  \bibinfo{volume}{34} (\bibinfo{year}{2009}), \bibinfo{pages}{405--430}.
\newblock


\bibitem[Mohamed et~al\mbox{.}(2020)]%
        {mohamed2020decolonial}
\bibfield{author}{\bibinfo{person}{Shakir Mohamed},
  \bibinfo{person}{Marie-Therese Png}, {and} \bibinfo{person}{William Isaac}.}
  \bibinfo{year}{2020}\natexlab{}.
\newblock \showarticletitle{Decolonial AI: Decolonial theory as sociotechnical
  foresight in artificial intelligence}.
\newblock \bibinfo{journal}{\emph{Philosophy \& Technology}}
  \bibinfo{volume}{33}, \bibinfo{number}{4} (\bibinfo{year}{2020}),
  \bibinfo{pages}{659--684}.
\newblock


\bibitem[Moitra et~al\mbox{.}(2022)]%
        {moitra2022ai}
\bibfield{author}{\bibinfo{person}{Aparna Moitra}, \bibinfo{person}{Dennis
  Wagenaar}, \bibinfo{person}{Manveer Kalirai}, \bibinfo{person}{Syed~Ishtiaque
  Ahmed}, {and} \bibinfo{person}{Robert Soden}.}
  \bibinfo{year}{2022}\natexlab{}.
\newblock \showarticletitle{AI and Disaster Risk: A Practitioner Perspective}.
\newblock \bibinfo{journal}{\emph{Proceedings of the ACM on Human-Computer
  Interaction}} \bibinfo{volume}{6}, \bibinfo{number}{CSCW2}
  (\bibinfo{year}{2022}), \bibinfo{pages}{1--20}.
\newblock


\bibitem[Moran(2010)]%
        {moran2010environmental}
\bibfield{author}{\bibinfo{person}{Emilio~F Moran}.}
  \bibinfo{year}{2010}\natexlab{}.
\newblock \bibinfo{booktitle}{\emph{Environmental social science:
  human-environment interactions and sustainability}}.
\newblock \bibinfo{publisher}{John Wiley \& Sons}.
\newblock


\bibitem[Musikanski et~al\mbox{.}(2018)]%
        {musikanski2018ieee}
\bibfield{author}{\bibinfo{person}{Laura Musikanski}, \bibinfo{person}{John
  Havens}, {and} \bibinfo{person}{Gregg Gunsch}.}
  \bibinfo{year}{2018}\natexlab{}.
\newblock \showarticletitle{IEEE P7010 Well-being Metrics Standard for
  Autonomous and Intelligent Systems}.
\newblock \bibinfo{journal}{\emph{IEEE, New York, NY, Tech. Rep}}
  (\bibinfo{year}{2018}).
\newblock


\bibitem[Nedzhvetskaya and Tan(2022)]%
        {nedzhvetskaya2022role}
\bibfield{author}{\bibinfo{person}{Nataliya Nedzhvetskaya} {and}
  \bibinfo{person}{JS Tan}.} \bibinfo{year}{2022}\natexlab{}.
\newblock \showarticletitle{The role of workers in AI ethics and governance}.
\newblock \bibinfo{journal}{\emph{The Oxford Handbook of AI Governance}}
  (\bibinfo{year}{2022}).
\newblock


\bibitem[Nkhata et~al\mbox{.}(2008)]%
        {nkhata2008resilient}
\bibfield{author}{\bibinfo{person}{Abraham~B Nkhata},
  \bibinfo{person}{Charles~M Breen}, {and} \bibinfo{person}{Wayne~A Freimund}.}
  \bibinfo{year}{2008}\natexlab{}.
\newblock \showarticletitle{Resilient social relationships and collaboration in
  the management of social--ecological systems}.
\newblock \bibinfo{journal}{\emph{Ecology and Society}} \bibinfo{volume}{13},
  \bibinfo{number}{1} (\bibinfo{year}{2008}).
\newblock


\bibitem[Noble(2018)]%
        {noble2018algorithms}
\bibfield{author}{\bibinfo{person}{Safiya~Umoja Noble}.}
  \bibinfo{year}{2018}\natexlab{}.
\newblock \showarticletitle{Algorithms of oppression}.
\newblock In \bibinfo{booktitle}{\emph{Algorithms of oppression}}.
  \bibinfo{publisher}{New York University Press}.
\newblock


\bibitem[Nost et~al\mbox{.}(021b)]%
        {nost2021visualizing}
\bibfield{author}{\bibinfo{person}{Eric Nost}, \bibinfo{person}{Gretchen
  Gehrke}, \bibinfo{person}{Grace Poudrier}, \bibinfo{person}{Aaron Lemelin},
  \bibinfo{person}{Marcy Beck}, \bibinfo{person}{Sara Wylie}, {and}
  \bibinfo{person}{Environmental Data \&~Governance Initiative}.}
  \bibinfo{year}{2021b}\natexlab{}.
\newblock \showarticletitle{Visualizing changes to US federal environmental
  agency websites}.
\newblock \bibinfo{journal}{\emph{PloS one}} \bibinfo{volume}{16},
  \bibinfo{number}{2} (\bibinfo{year}{2021b}), \bibinfo{pages}{e0246450}.
\newblock


\bibitem[Nost and Goldstein(021a)]%
        {eric2021}
\bibfield{author}{\bibinfo{person}{Eric Nost} {and} \bibinfo{person}{Jenny
  Goldstein}.} \bibinfo{year}{2021a}\natexlab{}.
\newblock \showarticletitle{A political ecology of data}.
\newblock \bibinfo{journal}{\emph{Environment and Planning E: Nature and
  Space}} (\bibinfo{date}{09} \bibinfo{year}{2021a}),
  \bibinfo{pages}{251484862110435}.
\newblock
\urldef\tempurl%
\url{https://doi.org/10.1177/25148486211043503}
\showDOI{\tempurl}


\bibitem[Ogbonnaya-Ogburu et~al\mbox{.}(2020)]%
        {ogbonnaya2020critical}
\bibfield{author}{\bibinfo{person}{Ihudiya~Finda Ogbonnaya-Ogburu},
  \bibinfo{person}{Angela~DR Smith}, \bibinfo{person}{Alexandra To}, {and}
  \bibinfo{person}{Kentaro Toyama}.} \bibinfo{year}{2020}\natexlab{}.
\newblock \showarticletitle{Critical race theory for HCI}. In
  \bibinfo{booktitle}{\emph{Proceedings of the 2020 CHI conference on human
  factors in computing systems}}. \bibinfo{pages}{1--16}.
\newblock


\bibitem[Orlikowski(2007)]%
        {doi:10.1177/0170840607081138}
\bibfield{author}{\bibinfo{person}{Wanda~J. Orlikowski}.}
  \bibinfo{year}{2007}\natexlab{}.
\newblock \showarticletitle{Sociomaterial Practices: Exploring Technology at
  Work}.
\newblock \bibinfo{journal}{\emph{Organization Studies}} \bibinfo{volume}{28},
  \bibinfo{number}{9} (\bibinfo{year}{2007}), \bibinfo{pages}{1435--1448}.
\newblock
\urldef\tempurl%
\url{https://doi.org/10.1177/0170840607081138}
\showDOI{\tempurl}


\bibitem[Ostrom(2007)]%
        {ostrom2007diagnostic}
\bibfield{author}{\bibinfo{person}{Elinor Ostrom}.}
  \bibinfo{year}{2007}\natexlab{}.
\newblock \showarticletitle{A diagnostic approach for going beyond panaceas}.
\newblock \bibinfo{journal}{\emph{Proceedings of the national Academy of
  sciences}} \bibinfo{volume}{104}, \bibinfo{number}{39}
  (\bibinfo{year}{2007}), \bibinfo{pages}{15181--15187}.
\newblock


\bibitem[Pasquinelli and Joler(2020)]%
        {pasquinelli2020nooscope}
\bibfield{author}{\bibinfo{person}{Matteo Pasquinelli} {and}
  \bibinfo{person}{Vladan Joler}.} \bibinfo{year}{2020}\natexlab{}.
\newblock \showarticletitle{The Nooscope manifested: AI as instrument of
  knowledge extractivism}.
\newblock \bibinfo{journal}{\emph{AI \& society}} (\bibinfo{year}{2020}),
  \bibinfo{pages}{1--18}.
\newblock


\bibitem[Pellow(2000)]%
        {pellow2000environmental}
\bibfield{author}{\bibinfo{person}{David~N Pellow}.}
  \bibinfo{year}{2000}\natexlab{}.
\newblock \showarticletitle{Environmental inequality formation: Toward a theory
  of environmental injustice}.
\newblock \bibinfo{journal}{\emph{American behavioral scientist}}
  \bibinfo{volume}{43}, \bibinfo{number}{4} (\bibinfo{year}{2000}),
  \bibinfo{pages}{581--601}.
\newblock


\bibitem[Pineda-Pinto et~al\mbox{.}(2021)]%
        {pineda2021examining}
\bibfield{author}{\bibinfo{person}{Melissa Pineda-Pinto},
  \bibinfo{person}{Pablo Herreros-Cantis}, \bibinfo{person}{Timon McPhearson},
  \bibinfo{person}{Niki Frantzeskaki}, \bibinfo{person}{Jing Wang}, {and}
  \bibinfo{person}{Weiqi Zhou}.} \bibinfo{year}{2021}\natexlab{}.
\newblock \showarticletitle{Examining ecological justice within the
  social-ecological-technological system of New York City, USA}.
\newblock \bibinfo{journal}{\emph{Landscape and Urban Planning}}
  \bibinfo{volume}{215} (\bibinfo{year}{2021}), \bibinfo{pages}{104228}.
\newblock


\bibitem[Png(2022)]%
        {png2022tensions}
\bibfield{author}{\bibinfo{person}{Marie-Therese Png}.}
  \bibinfo{year}{2022}\natexlab{}.
\newblock \showarticletitle{At the Tensions of South and North: Critical Roles
  of Global South Stakeholders in AI Governance}. In
  \bibinfo{booktitle}{\emph{2022 ACM Conference on Fairness, Accountability,
  and Transparency}}. \bibinfo{pages}{1434--1445}.
\newblock


\bibitem[Raji and Buolamwini(2019)]%
        {raji2019actionable}
\bibfield{author}{\bibinfo{person}{Inioluwa~Deborah Raji} {and}
  \bibinfo{person}{Joy Buolamwini}.} \bibinfo{year}{2019}\natexlab{}.
\newblock \showarticletitle{Actionable auditing: Investigating the impact of
  publicly naming biased performance results of commercial ai products}. In
  \bibinfo{booktitle}{\emph{Proceedings of the 2019 AAAI/ACM Conference on AI,
  Ethics, and Society}}. \bibinfo{pages}{429--435}.
\newblock


\bibitem[Raji et~al\mbox{.}(2020)]%
        {raji2020closing}
\bibfield{author}{\bibinfo{person}{Inioluwa~Deborah Raji},
  \bibinfo{person}{Andrew Smart}, \bibinfo{person}{Rebecca~N White},
  \bibinfo{person}{Margaret Mitchell}, \bibinfo{person}{Timnit Gebru},
  \bibinfo{person}{Ben Hutchinson}, \bibinfo{person}{Jamila Smith-Loud},
  \bibinfo{person}{Daniel Theron}, {and} \bibinfo{person}{Parker Barnes}.}
  \bibinfo{year}{2020}\natexlab{}.
\newblock \showarticletitle{Closing the AI accountability gap: Defining an
  end-to-end framework for internal algorithmic auditing}. In
  \bibinfo{booktitle}{\emph{Proceedings of the 2020 conference on fairness,
  accountability, and transparency}}. \bibinfo{pages}{33--44}.
\newblock


\bibitem[Rakova et~al\mbox{.}(2021)]%
        {rakova2021responsible}
\bibfield{author}{\bibinfo{person}{Bogdana Rakova}, \bibinfo{person}{Jingying
  Yang}, \bibinfo{person}{Henriette Cramer}, {and} \bibinfo{person}{Rumman
  Chowdhury}.} \bibinfo{year}{2021}\natexlab{}.
\newblock \showarticletitle{Where responsible AI meets reality: Practitioner
  perspectives on enablers for shifting organizational practices}.
\newblock \bibinfo{journal}{\emph{Proceedings of the ACM on Human-Computer
  Interaction}} \bibinfo{volume}{5}, \bibinfo{number}{CSCW1}
  (\bibinfo{year}{2021}), \bibinfo{pages}{1--23}.
\newblock


\bibitem[Ricaurte(2019)]%
        {ricaurte2019data}
\bibfield{author}{\bibinfo{person}{Paola Ricaurte}.}
  \bibinfo{year}{2019}\natexlab{}.
\newblock \showarticletitle{Data epistemologies, the coloniality of power, and
  resistance}.
\newblock \bibinfo{journal}{\emph{Television \& New Media}}
  \bibinfo{volume}{20}, \bibinfo{number}{4} (\bibinfo{year}{2019}),
  \bibinfo{pages}{350--365}.
\newblock


\bibitem[Rohde et~al\mbox{.}(2021)]%
        {rohde2021sustainability}
\bibfield{author}{\bibinfo{person}{Friederike Rohde}, \bibinfo{person}{Maike
  Gossen}, \bibinfo{person}{Josephin Wagner}, {and} \bibinfo{person}{Tilman
  Santarius}.} \bibinfo{year}{2021}\natexlab{}.
\newblock \showarticletitle{Sustainability challenges of Artificial
  Intelligence and Policy Implications}.
\newblock \bibinfo{journal}{\emph{{\"O}kologisches
  Wirtschaften-Fachzeitschrift}} \bibinfo{volume}{36}, \bibinfo{number}{O1}
  (\bibinfo{year}{2021}), \bibinfo{pages}{36--40}.
\newblock


\bibitem[Rolnick et~al\mbox{.}(2022)]%
        {rolnick2022tackling}
\bibfield{author}{\bibinfo{person}{David Rolnick}, \bibinfo{person}{Priya~L
  Donti}, \bibinfo{person}{Lynn~H Kaack}, \bibinfo{person}{Kelly Kochanski},
  \bibinfo{person}{Alexandre Lacoste}, \bibinfo{person}{Kris Sankaran},
  \bibinfo{person}{Andrew~Slavin Ross}, \bibinfo{person}{Nikola
  Milojevic-Dupont}, \bibinfo{person}{Natasha Jaques}, \bibinfo{person}{Anna
  Waldman-Brown}, {et~al\mbox{.}}} \bibinfo{year}{2022}\natexlab{}.
\newblock \showarticletitle{Tackling climate change with machine learning}.
\newblock \bibinfo{journal}{\emph{ACM Computing Surveys (CSUR)}}
  \bibinfo{volume}{55}, \bibinfo{number}{2} (\bibinfo{year}{2022}),
  \bibinfo{pages}{1--96}.
\newblock


\bibitem[Rozzi et~al\mbox{.}(2015)]%
        {rozzi2015earth}
\bibfield{author}{\bibinfo{person}{Ricardo Rozzi}, \bibinfo{person}{F~Stuart
  Chapin~III}, \bibinfo{person}{J~Baird Callicott}, \bibinfo{person}{Steward~TA
  Pickett}, \bibinfo{person}{Mary~E Power}, \bibinfo{person}{Juan~J Armesto},
  {and} \bibinfo{person}{Roy~H May~Jr}.} \bibinfo{year}{2015}\natexlab{}.
\newblock \bibinfo{booktitle}{\emph{Earth stewardship: linking ecology and
  ethics in theory and practice}}. Vol.~\bibinfo{volume}{2}.
\newblock \bibinfo{publisher}{Springer}.
\newblock


\bibitem[Schwartz et~al\mbox{.}(2020)]%
        {schwartz2020green}
\bibfield{author}{\bibinfo{person}{Roy Schwartz}, \bibinfo{person}{Jesse
  Dodge}, \bibinfo{person}{Noah~A Smith}, {and} \bibinfo{person}{Oren
  Etzioni}.} \bibinfo{year}{2020}\natexlab{}.
\newblock \showarticletitle{Green ai}.
\newblock \bibinfo{journal}{\emph{Commun. ACM}} \bibinfo{volume}{63},
  \bibinfo{number}{12} (\bibinfo{year}{2020}), \bibinfo{pages}{54--63}.
\newblock


\bibitem[Seaver(2019)]%
        {seaver2019captivating}
\bibfield{author}{\bibinfo{person}{Nick Seaver}.}
  \bibinfo{year}{2019}\natexlab{}.
\newblock \showarticletitle{Captivating algorithms: Recommender systems as
  traps}.
\newblock \bibinfo{journal}{\emph{Journal of Material Culture}}
  \bibinfo{volume}{24}, \bibinfo{number}{4} (\bibinfo{year}{2019}),
  \bibinfo{pages}{421--436}.
\newblock


\bibitem[Selbst et~al\mbox{.}(2019)]%
        {selbst2019fairness}
\bibfield{author}{\bibinfo{person}{Andrew~D Selbst}, \bibinfo{person}{Danah
  Boyd}, \bibinfo{person}{Sorelle~A Friedler}, \bibinfo{person}{Suresh
  Venkatasubramanian}, {and} \bibinfo{person}{Janet Vertesi}.}
  \bibinfo{year}{2019}\natexlab{}.
\newblock \showarticletitle{Fairness and abstraction in sociotechnical
  systems}. In \bibinfo{booktitle}{\emph{Proceedings of the conference on
  fairness, accountability, and transparency}}. \bibinfo{pages}{59--68}.
\newblock


\bibitem[Shen et~al\mbox{.}(2021)]%
        {shen2021everyday}
\bibfield{author}{\bibinfo{person}{Hong Shen}, \bibinfo{person}{Alicia DeVos},
  \bibinfo{person}{Motahhare Eslami}, {and} \bibinfo{person}{Kenneth
  Holstein}.} \bibinfo{year}{2021}\natexlab{}.
\newblock \showarticletitle{Everyday algorithm auditing: Understanding the
  power of everyday users in surfacing harmful algorithmic behaviors}.
\newblock \bibinfo{journal}{\emph{Proceedings of the ACM on Human-Computer
  Interaction}} \bibinfo{volume}{5}, \bibinfo{number}{CSCW2}
  (\bibinfo{year}{2021}), \bibinfo{pages}{1--29}.
\newblock


\bibitem[Sloane(2019)]%
        {sloane2019inequality}
\bibfield{author}{\bibinfo{person}{Mona Sloane}.}
  \bibinfo{year}{2019}\natexlab{}.
\newblock \showarticletitle{Inequality is the name of the game: thoughts on the
  emerging field of technology, ethics and social justice}. In
  \bibinfo{booktitle}{\emph{Weizenbaum Conference}}. DEU, \bibinfo{pages}{9}.
\newblock


\bibitem[Sloane and Moss(2019)]%
        {sloane2019ai}
\bibfield{author}{\bibinfo{person}{Mona Sloane} {and} \bibinfo{person}{Emanuel
  Moss}.} \bibinfo{year}{2019}\natexlab{}.
\newblock \showarticletitle{AI’s social sciences deficit}.
\newblock \bibinfo{journal}{\emph{Nature Machine Intelligence}}
  \bibinfo{volume}{1}, \bibinfo{number}{8} (\bibinfo{year}{2019}),
  \bibinfo{pages}{330--331}.
\newblock


\bibitem[Sloane et~al\mbox{.}(2020)]%
        {sloane2020participation}
\bibfield{author}{\bibinfo{person}{Mona Sloane}, \bibinfo{person}{Emanuel
  Moss}, \bibinfo{person}{Olaitan Awomolo}, {and} \bibinfo{person}{Laura
  Forlano}.} \bibinfo{year}{2020}\natexlab{}.
\newblock \showarticletitle{Participation is not a design fix for machine
  learning}.
\newblock \bibinfo{journal}{\emph{arXiv preprint arXiv:2007.02423}}
  (\bibinfo{year}{2020}).
\newblock


\bibitem[Smith and Stirling(2010)]%
        {smith2010politics}
\bibfield{author}{\bibinfo{person}{Adrian Smith} {and} \bibinfo{person}{Andy
  Stirling}.} \bibinfo{year}{2010}\natexlab{}.
\newblock \showarticletitle{The politics of social-ecological resilience and
  sustainable socio-technical transitions}.
\newblock \bibinfo{journal}{\emph{Ecology and society}} \bibinfo{volume}{15},
  \bibinfo{number}{1} (\bibinfo{year}{2010}).
\newblock


\bibitem[Star(1990)]%
        {doi:10.1111/j.1467-954X.1990.tb03347.x}
\bibfield{author}{\bibinfo{person}{Susan~Leigh Star}.}
  \bibinfo{year}{1990}\natexlab{}.
\newblock \showarticletitle{Power, Technology and the Phenomenology of
  Conventions: On being Allergic to Onions}.
\newblock \bibinfo{journal}{\emph{The Sociological Review}}
  \bibinfo{volume}{38}, \bibinfo{number}{1\_suppl} (\bibinfo{year}{1990}),
  \bibinfo{pages}{26--56}.
\newblock
\urldef\tempurl%
\url{https://doi.org/10.1111/j.1467-954X.1990.tb03347.x}
\showDOI{\tempurl}


\bibitem[Stark(2019)]%
        {stark2019facial}
\bibfield{author}{\bibinfo{person}{Luke Stark}.}
  \bibinfo{year}{2019}\natexlab{}.
\newblock \showarticletitle{Facial recognition is the plutonium of AI}.
\newblock \bibinfo{journal}{\emph{XRDS: Crossroads, The ACM Magazine for
  Students}} \bibinfo{volume}{25}, \bibinfo{number}{3} (\bibinfo{year}{2019}),
  \bibinfo{pages}{50--55}.
\newblock


\bibitem[Stark and Hoffmann(2019)]%
        {stark2019data}
\bibfield{author}{\bibinfo{person}{Luke Stark} {and}
  \bibinfo{person}{Anna~Lauren Hoffmann}.} \bibinfo{year}{2019}\natexlab{}.
\newblock \showarticletitle{Data is the new what? Popular metaphors \&
  professional ethics in emerging data culture}.
\newblock  (\bibinfo{year}{2019}).
\newblock


\bibitem[Stibbe(2015)]%
        {stibbe2015ecolinguistics}
\bibfield{author}{\bibinfo{person}{Arran Stibbe}.}
  \bibinfo{year}{2015}\natexlab{}.
\newblock \bibinfo{booktitle}{\emph{Ecolinguistics: Language, ecology and the
  stories we live by}}.
\newblock \bibinfo{publisher}{Routledge}.
\newblock


\bibitem[Stirling(2008)]%
        {stirling2008opening}
\bibfield{author}{\bibinfo{person}{Andy Stirling}.}
  \bibinfo{year}{2008}\natexlab{}.
\newblock \showarticletitle{“Opening up” and “closing down” power,
  participation, and pluralism in the social appraisal of technology}.
\newblock \bibinfo{journal}{\emph{Science, Technology, \& Human Values}}
  \bibinfo{volume}{33}, \bibinfo{number}{2} (\bibinfo{year}{2008}),
  \bibinfo{pages}{262--294}.
\newblock


\bibitem[Strubell et~al\mbox{.}(2019)]%
        {strubell2019energy}
\bibfield{author}{\bibinfo{person}{Emma Strubell}, \bibinfo{person}{Ananya
  Ganesh}, {and} \bibinfo{person}{Andrew McCallum}.}
  \bibinfo{year}{2019}\natexlab{}.
\newblock \showarticletitle{Energy and policy considerations for deep learning
  in NLP}.
\newblock \bibinfo{journal}{\emph{arXiv preprint arXiv:1906.02243}}
  (\bibinfo{year}{2019}).
\newblock


\bibitem[Taiuru(2021)]%
        {maori2021}
\bibfield{author}{\bibinfo{person}{Karaitiana Taiuru}.}
  \bibinfo{year}{2021}\natexlab{}.
\newblock \bibinfo{title}{Māori Data Sovereignty Licences}.
\newblock
\newblock
\urldef\tempurl%
\url{https://www.taiuru.maori.nz/maori-data-sovereignty-licences/}
\showURL{%
\tempurl}


\bibitem[Taylor(2000)]%
        {taylor2000rise}
\bibfield{author}{\bibinfo{person}{Dorceta~E Taylor}.}
  \bibinfo{year}{2000}\natexlab{}.
\newblock \showarticletitle{The rise of the environmental justice paradigm:
  Injustice framing and the social construction of environmental discourses}.
\newblock \bibinfo{journal}{\emph{American behavioral scientist}}
  \bibinfo{volume}{43}, \bibinfo{number}{4} (\bibinfo{year}{2000}),
  \bibinfo{pages}{508--580}.
\newblock


\bibitem[Temper(2019)]%
        {temper2019blocking}
\bibfield{author}{\bibinfo{person}{Leah Temper}.}
  \bibinfo{year}{2019}\natexlab{}.
\newblock \showarticletitle{Blocking pipelines, unsettling environmental
  justice: from rights of nature to responsibility to territory}.
\newblock \bibinfo{journal}{\emph{Local Environment}} \bibinfo{volume}{24},
  \bibinfo{number}{2} (\bibinfo{year}{2019}), \bibinfo{pages}{94--112}.
\newblock


\bibitem[Thatcher et~al\mbox{.}(2016)]%
        {thatcher2016data}
\bibfield{author}{\bibinfo{person}{Jim Thatcher}, \bibinfo{person}{David
  O’Sullivan}, {and} \bibinfo{person}{Dillon Mahmoudi}.}
  \bibinfo{year}{2016}\natexlab{}.
\newblock \showarticletitle{Data colonialism through accumulation by
  dispossession: New metaphors for daily data}.
\newblock \bibinfo{journal}{\emph{Environment and Planning D: Society and
  Space}} \bibinfo{volume}{34}, \bibinfo{number}{6} (\bibinfo{year}{2016}),
  \bibinfo{pages}{990--1006}.
\newblock


\bibitem[Tobin(1997)]%
        {tobin1997natural}
\bibfield{author}{\bibinfo{person}{Graham~A Tobin}.}
  \bibinfo{year}{1997}\natexlab{}.
\newblock \bibinfo{booktitle}{\emph{Natural hazards: explanation and
  integration}}.
\newblock \bibinfo{publisher}{Guilford Press}.
\newblock


\bibitem[Turner et~al\mbox{.}(2003)]%
        {turner2003framework}
\bibfield{author}{\bibinfo{person}{Billie~L Turner}, \bibinfo{person}{Roger~E
  Kasperson}, \bibinfo{person}{Pamela~A Matson}, \bibinfo{person}{James~J
  McCarthy}, \bibinfo{person}{Robert~W Corell}, \bibinfo{person}{Lindsey
  Christensen}, \bibinfo{person}{Noelle Eckley}, \bibinfo{person}{Jeanne~X
  Kasperson}, \bibinfo{person}{Amy Luers}, \bibinfo{person}{Marybeth~L
  Martello}, {et~al\mbox{.}}} \bibinfo{year}{2003}\natexlab{}.
\newblock \showarticletitle{A framework for vulnerability analysis in
  sustainability science}.
\newblock \bibinfo{journal}{\emph{Proceedings of the national academy of
  sciences}} \bibinfo{volume}{100}, \bibinfo{number}{14}
  (\bibinfo{year}{2003}), \bibinfo{pages}{8074--8079}.
\newblock


\bibitem[Umbrello(2021)]%
        {umbrello2021ecological}
\bibfield{author}{\bibinfo{person}{Steven Umbrello}.}
  \bibinfo{year}{2021}\natexlab{}.
\newblock \showarticletitle{The Ecological Turn in Design: Adopting a
  Posthumanist Ethics to Inform Value Sensitive Design}.
\newblock \bibinfo{journal}{\emph{Philosophies}} \bibinfo{volume}{6},
  \bibinfo{number}{2} (\bibinfo{year}{2021}), \bibinfo{pages}{29}.
\newblock


\bibitem[van Wynsberghe(2021)]%
        {van2021sustainable}
\bibfield{author}{\bibinfo{person}{Aimee van Wynsberghe}.}
  \bibinfo{year}{2021}\natexlab{}.
\newblock \showarticletitle{Sustainable AI: AI for sustainability and the
  sustainability of AI}.
\newblock \bibinfo{journal}{\emph{AI and Ethics}} \bibinfo{volume}{1},
  \bibinfo{number}{3} (\bibinfo{year}{2021}), \bibinfo{pages}{213--218}.
\newblock


\bibitem[Viljoen(2021)]%
        {viljoen2021relational}
\bibfield{author}{\bibinfo{person}{Salome Viljoen}.}
  \bibinfo{year}{2021}\natexlab{}.
\newblock \showarticletitle{A relational theory of data governance}.
\newblock \bibinfo{journal}{\emph{Yale LJ}}  \bibinfo{volume}{131}
  (\bibinfo{year}{2021}), \bibinfo{pages}{573}.
\newblock


\bibitem[Wald et~al\mbox{.}(2019)]%
        {wald2019latinx}
\bibfield{author}{\bibinfo{person}{Sarah~D Wald}, \bibinfo{person}{David~J
  V{\'a}zquez}, \bibinfo{person}{Priscilla~Solis Ybarra},
  \bibinfo{person}{Sarah~Jaquette Ray}, \bibinfo{person}{Laura Pulido}, {and}
  \bibinfo{person}{Stacy Alaimo}.} \bibinfo{year}{2019}\natexlab{}.
\newblock \bibinfo{booktitle}{\emph{Latinx environmentalisms: Place, justice,
  and the decolonial}}.
\newblock \bibinfo{publisher}{Temple University Press}.
\newblock


\bibitem[Walker(2009)]%
        {walker2009beyond}
\bibfield{author}{\bibinfo{person}{Gordon Walker}.}
  \bibinfo{year}{2009}\natexlab{}.
\newblock \showarticletitle{Beyond distribution and proximity: exploring the
  multiple spatialities of environmental justice}.
\newblock \bibinfo{journal}{\emph{Antipode}} \bibinfo{volume}{41},
  \bibinfo{number}{4} (\bibinfo{year}{2009}), \bibinfo{pages}{614--636}.
\newblock


\bibitem[West et~al\mbox{.}(2020)]%
        {west2020relational}
\bibfield{author}{\bibinfo{person}{Simon West}, \bibinfo{person}{L~Jamila
  Haider}, \bibinfo{person}{Sanna St{\aa}lhammar}, {and}
  \bibinfo{person}{Stephen Woroniecki}.} \bibinfo{year}{2020}\natexlab{}.
\newblock \showarticletitle{A relational turn for sustainability science?
  Relational thinking, leverage points and transformations}.
\newblock \bibinfo{journal}{\emph{Ecosystems and People}} \bibinfo{volume}{16},
  \bibinfo{number}{1} (\bibinfo{year}{2020}), \bibinfo{pages}{304--325}.
\newblock


\bibitem[Westra and Lawson(2001)]%
        {westra2001faces}
\bibfield{author}{\bibinfo{person}{Laura Westra} {and} \bibinfo{person}{Bill
  Lawson}.} \bibinfo{year}{2001}\natexlab{}.
\newblock \bibinfo{booktitle}{\emph{Faces of environmental racism: Confronting
  issues of global justice}}.
\newblock \bibinfo{publisher}{Rowman \& Littlefield Publishers}.
\newblock


\bibitem[Whittaker(2021)]%
        {whittaker_steep_2021}
\bibfield{author}{\bibinfo{person}{Meredith Whittaker}.}
  \bibinfo{year}{2021}\natexlab{}.
\newblock \showarticletitle{The steep cost of capture}.
\newblock \bibinfo{journal}{\emph{Interactions}} \bibinfo{volume}{28},
  \bibinfo{number}{6} (\bibinfo{year}{2021}), \bibinfo{pages}{50--55}.
\newblock
\newblock
\shownote{Publisher: ACM New York, NY, USA}.


\bibitem[Widder and Nafus(2022)]%
        {widder2022dislocated}
\bibfield{author}{\bibinfo{person}{David~Gray Widder} {and}
  \bibinfo{person}{Dawn Nafus}.} \bibinfo{year}{2022}\natexlab{}.
\newblock \showarticletitle{Dislocated Accountabilities in the AI Supply Chain:
  Modularity and Developers' Notions of Responsibility}.
\newblock \bibinfo{journal}{\emph{arXiv preprint arXiv:2209.09780}}
  (\bibinfo{year}{2022}).
\newblock


\bibitem[Xu et~al\mbox{.}(2021)]%
        {xu2021survey}
\bibfield{author}{\bibinfo{person}{Jingjing Xu}, \bibinfo{person}{Wangchunshu
  Zhou}, \bibinfo{person}{Zhiyi Fu}, \bibinfo{person}{Hao Zhou}, {and}
  \bibinfo{person}{Lei Li}.} \bibinfo{year}{2021}\natexlab{}.
\newblock \showarticletitle{A Survey on Green Deep Learning}.
\newblock \bibinfo{journal}{\emph{arXiv preprint arXiv:2111.05193}}
  (\bibinfo{year}{2021}).
\newblock


\bibitem[Young et~al\mbox{.}(2022)]%
        {young_confronting_2022}
\bibfield{author}{\bibinfo{person}{Meg Young}, \bibinfo{person}{Michael
  Katell}, {and} \bibinfo{person}{P.M. Krafft}.}
  \bibinfo{year}{2022}\natexlab{}.
\newblock \showarticletitle{Confronting {Power} and {Corporate} {Capture} at
  the {FAccT} {Conference}}. In \bibinfo{booktitle}{\emph{2022 {ACM}
  {Conference} on {Fairness}, {Accountability}, and {Transparency}}}
  \emph{(\bibinfo{series}{{FAccT} '22})}. \bibinfo{publisher}{Association for
  Computing Machinery}, \bibinfo{address}{New York, NY, USA},
  \bibinfo{pages}{1375--1386}.
\newblock
\showISBNx{978-1-4503-9352-2}
\urldef\tempurl%
\url{https://doi.org/10.1145/3531146.3533194}
\showDOI{\tempurl}


\bibitem[Ziesche(2021)]%
        {ziesche2021ai}
\bibfield{author}{\bibinfo{person}{Soenke Ziesche}.}
  \bibinfo{year}{2021}\natexlab{}.
\newblock \showarticletitle{AI Ethics and value alignment for nonhuman
  animals}.
\newblock \bibinfo{journal}{\emph{Philosophies}} \bibinfo{volume}{6},
  \bibinfo{number}{2} (\bibinfo{year}{2021}), \bibinfo{pages}{31}.
\newblock


\end{thebibliography}
